\documentclass[preprint,nofootinbib,aps,pra,a4paper,preprintnumbers,showpacs,showkeys]{revtex4-1}

\usepackage{amssymb}
\usepackage{amsmath}
\usepackage{graphicx}
\usepackage{dcolumn}
\usepackage{bm}
\usepackage{epsfig}
\usepackage{hyperref}

\def\figdir{}

\newcommand\beq{\begin{eqnarray}}
\newcommand\eeq{\end{eqnarray}}

\newcommand\figwidth{.40\textwidth}
\newcommand\Eq[1]{Eq.~\ref{eq:#1}}
\newcommand\Fig[1]{Fig.~\ref{fig:#1}}
\newcommand\App[1]{Appendix~\ref{app:#1}}
\newcommand\Sec[1]{Sec.~\ref{sec:#1}}
\newcommand\Tab[1]{Table~\ref{tab:#1}}

\newcommand\bfn{\mathbf n}

\newcommand\bfe{\mathbf e}

\newcommand\calA{\mathcal A}
\newcommand\calB{\mathcal B}

\newcommand\calC{\mathcal C}
\newcommand\calD{\mathcal D}

\newcommand\calE{\mathcal E}
\newcommand\calL{\mathcal L}
\newcommand\calO{\mathcal O}

\newcommand\calV{\mathcal V}

\begin{document}

\preprint{RIKEN-QHP-65}

\title{Numerical study of unitary fermions in one spatial dimension}

\author{Michael G. Endres}
\email{endres@riken.jp}

\affiliation{Theoretical Research Division, RIKEN Nishina Center, Wako, Saitama 351-0198, Japan}


\date{\today}
 
\begin{abstract}
I perform lattice Monte Carlo studies of universal four-component fermion systems in one spatial dimension.
Continuum few-body observables (i.e., ground state energies and integrated contact densities) are determined for both unpolarized and polarized systems of up to eight fermions confined to a harmonic trap.
Estimates of the continuum energies for four and five trapped fermions show agreement with exact analytic calculations to within approximately one percent statistical uncertainties.
Continuum many-body observables are determined for unpolarized systems of up to 88 fermions confined to a finite box, and 56 fermions confined to a harmonic trap.
Results are reported for universal quantities such as the Bertsch parameter, defined as the energy of the untrapped many-body system in units of the corresponding free-gas energy, and its subleading correction at large but finite scattering length.
Two independent estimates of these quantities are obtained from thermodynamic limit extrapolations of continuum extrapolated observables.
A third estimate of the Bertsch parameter is obtained by combining estimates of the untrapped and trapped integrated contact densities with additional theoretical input from a calculation based on Thomas-Fermi theory.
All estimates of the Bertsch parameter and its subleading correction are found to be consistent to within approximately one percent statistical uncertainties.
Finally, the continuum restoration of virial theorems is verified for both few- and many-body systems confined to a trap.
\end{abstract}

\maketitle

\section{Introduction}
\label{sec:introduction}

Universal Fermi gases have garnered wide-spread attention following their recent realization in ultracold atom experiments.
Perhaps the most physically interesting example of such a system is the unitary Fermi gas in three spatial dimensions.
This system comprises a dilute mixture of spin-1/2 (i.e., two-component) fermions interacting via a short-range interparticle potential tuned to produce an infinite two-particle scattering length.
It is universal in the sense that the physical properties of the system are independent of the short-distance character of the interaction.
As such, unitary fermions are not only relevant for describing ultracold atoms, but also a variety of physical systems considered in other disciplines as well.

Early on, the unitary Fermi gas was proposed as an idealized model for describing dilute neutron matter in neutron stars \cite{1995ARNPS..45..429P}.
The system had later been realized in ultracold atom experiments by exploiting properties of a Feshbach resonance \cite{O'Hara13122002,PhysRevLett.93.050401,2003Natur.424...47R,PhysRevLett.91.080406,PhysRevLett.89.203201}.
More recently, unitary fermions have gained theoretical interest as an example of a nonrelativistic conformal field theory. 
A greater experimental and theoretical understanding of this strongly interacting and nonperturbative system has revealed that despite their simplicity, unitary fermions possess many rich and rather surprising physical properties.

The unitary Fermi gas is simple in the sense that it is characterized by a single physical scale, the density $\rho$.
From purely dimensional considerations, the energy density of the interacting system must be given by
\begin{eqnarray}
\calE(\rho) = \xi \calE_{0}(\rho) \ ,
\label{eq:untrapped_energy}
\end{eqnarray}
where $\calE_{0}(\rho)$ is the energy density for free fermions, and the dimensionless constant of proportionality $\xi$ is a nonperturbative universal quantity known as the Bertsch parameter \cite{baker2000mbx}.
A calculation based on density-functional theory \cite{PhysRevA.72.041603} suggests that for the same system confined to a harmonic trap, the energy is given by
\begin{eqnarray}
E^{osc}(Q) = \sqrt{\xi} E^{osc}_{0}(Q) \ ,
\label{eq:trapped_energy}
\end{eqnarray}
where $E^{osc}_{0}(Q)$ is the energy of the corresponding noninteracting system in the limit of large total fermion number $Q$.
The parameter $\xi$ appearing as the square of the constant of proportionality in \Eq{trapped_energy} is the same as that in \Eq{untrapped_energy}.
This nontrivial relation, including subleading corrections due to a finite fermion number, was independently confirmed from a calculation based on a general coordinate invariant effective field theory description of the system \cite{Son:2005rv,2009AnPhy.324.1136M}.

During the past decade, substantial effort has been devoted toward determining properties of the unitary Fermi gas using a variety of analytical, numerical and experimental means (see, e.g., \cite{PhysRevA.87.023615} and references therein for a historical summary of Bertsch parameter determinations made by various methods).
Although analytical calculations have become quite sophisticated, they generically possess unquantifiable systematic errors due to the nonperturbative nature of the problem.
Numerical simulations on the other hand, while in principle provide an exact avenue toward nonperturbative results, are often hampered by signal/noise and sign problems.
Presently the domain of utility for such simulations have been largely confined to the $SU(2)$ symmetric point, where population densities and masses associated with each spin degree of freedom are equal, and its vicinity.
Despite these limitations, however, numerical simulations have been used quite successfully to obtain reliable quantitative results where applicable. 
For example, the current best estimate for the Bertsch parameter from lattice Monte Carlo studies of an untrapped Fermi gas have yielded $\xi=0.372(5)$ \cite{PhysRevA.84.061602}, whereas the most accurately determined value for the Bertsch parameter based on experiment is $\xi=0.376(4)$ \cite{Ku03022012}.

More recently, interest has turned toward Fermi gases in lower and mixed dimensions, which might in principle be created in ultracold atom experiments using strong optical lattices.
It had been argued that such systems, when finely tuned, can exhibit scale-invariance and have universal properties very much analogous to the conventional three-dimensional unitary Fermi gas \cite{PhysRevLett.101.170401}.
One particularly interesting example in one spatial dimension is a dilute four-component gas of nonrelativistic fermions interacting via an attractive short-range four-particle potential.
In the vicinity of a four-body resonance, Nishida and Son had demonstrated an exact mapping between the four-body problem for this system and the three-dimensional two-body problem for spin-1/2 fermions in the unitary regime \cite{Nishida:2009pg}.
It was argued that in this regime, the long-distance properties of the four-particle interaction can be characterized by a single length-scale ($a$).
This scale may be regarded as a one-dimensional analog of the conventional scattering length which characterizes a short-range two-particle interaction in three-dimensions.
At resonance, the scattering length diverges and the one-dimensional system becomes universal in the same sense as its three-dimensional counterpart: properties of the system become independent of the details of the interaction.
Because the scale and conformal invariance of the one-dimensional system is realized at a four-body resonance, its constituents are often referred to as ``unitary fermions'' in analogy with the corresponding three-dimensional system.

Although the mapping does not extend to the many-body problem, it was nevertheless demonstrated in \cite{Nishida:2009pg} that the zero-temperature few- and many-body properties of the one-dimensional Fermi gas are in many ways qualitatively identical to those of the three-dimensional spin-1/2 Fermi gas.
For example, in the strong coupling limit, corresponding to a small positive scattering length, the one-dimensional system becomes a dilute gas of tightly bound tetramers, and may be viewed as a one-dimensional analog of a Bose-Einstein condensate (BEC) in three dimensions.
In the weak coupling limit, corresponding to a small negative scattering length, the system exhibits properties that are strikingly similar to those of the BCS regime in three dimensions (e.g., a gap spectrum that is exponentially small in the inverse-scattering length).
These two regimes are continuously connected by varying the scattering length, and in the limit of infinite scattering length, the one-dimensional system exhibits properties similar to those of the BEC-BCS cross-over.

In the unitary limit, the one dimensional Fermi gas is characterized by a single scale, the density of the system.
By dimensional analysis, it follows that the energy density of the untrapped system must obey \Eq{untrapped_energy}, where the free gas energy density is given by
\begin{eqnarray}
\calE_0(\rho) = \frac{\pi^2 \rho^3}{96m}\ ,
\label{eq:free_gas_energy}
\end{eqnarray}
and the constant of proportionality $\xi$ is the one-dimensional analog of the three-dimensional Bertsch parameter.
Similarly, a calculation based on Thomas-Fermi theory along the lines of \cite{PhysRevA.72.041603} shows that the one-dimensional Fermi gas confined to a harmonic trap obeys \Eq{trapped_energy} at unitarity, where
\begin{eqnarray}
E^{osc}_0(Q) = \frac{1}{8} Q^2 \omega\ ,
\label{eq:free_trapped_gas_energy}
\end{eqnarray}
and $\omega$ is the characteristic frequency of the harmonic potential\footnote{Throughout I work in units where $\hbar = 1$.} (see \App{thomas-fermi} for details).
The Bertsch parameter associated with the four-component Fermi gas need not be the same as that of the two-component Fermi gas, and at present there are no known theoretical arguments that establish that they are.
Numerical evidence reported in a companion paper, however, suggests that the Bertsch parameter for these two systems are in fact equal to within approximately one percent statistical uncertainties \cite{PhysRevLett.109.250403}.
The results hint at a possible duality between the one- and three-dimensional Fermi gases extending beyond the few-body problem, and in a search for additional clues provides a partial motivation for further study of the one-dimensional system.

At finite scattering length the energy density of the untrapped one-dimensional unitary Fermi gas, normalized by the the free-gas energy density, can be described by a universal Bertsch function, $\Xi(y)$, which depends solely on the dimensionless quantity $y = (k_F a)^{-1}$, where $k_F=\pi\rho/4$ is the Fermi momentum for free fermions.
This function was shown to behave as \cite{Nishida:2009pg}
\begin{eqnarray}
\Xi(y) =  \left\{
\begin{array}{ll}
1 + \frac{6}{\pi^2 y} + \ldots        & \qquad y\ll -1 \\
\xi - \zeta y + \ldots   & \qquad |y| \ll 1\\
-\frac{3}{4}y^2 + \ldots  & \qquad y \gg 1\\
\end{array}
\right.
\label{eq:bertsch_function}
\end{eqnarray}
where the Bertsch parameter $\xi$ was previously discussed, and the slope $\zeta$ of $\Xi$ at unitarity is a second unknown nonperturbative constant.
The slope of the Bertsch function in the unitary limit can be expressed as
\begin{eqnarray}
\zeta =  \frac{3}{2\pi} \frac{\calC}{\rho k_F}\ ,
\end{eqnarray}
where the quantity
\begin{eqnarray}
\calC \equiv -(4\pi m) \left. \frac{d \calE}{d a^{-1}} \right|_{a=\infty} \ .
\label{eq:contact_definition}
\end{eqnarray}
is known as the contact density, evaluated in the unitary limit.
The contact density is a well-defined physical quantity for both one- and three- dimensional theories (although they are not necessarily equal), and plays a fundamental role in various universal (Tan) relations \cite{2008AnPhy.323.2952T,2008AnPhy.323.2971T,2008AnPhy.323.2987T,PhysRevLett.100.205301,2008PhRvA..78e3606B}.
Furthermore, it may be defined for both few- and many-body systems, at zero- and finite-temperature, and even away from unitarity.

Other common properties that unitary fermions in one and three dimensions share follow from the fact that the Hamiltonians for both systems are invariant under symmetry transformations generated by the Schr\"{o}dinger algebra \cite{PhysRevD.5.377,Niederer:1972zz}.
The theoretical implications of these symmetries have been explored in great detail \cite{PhysRevD.76.086004,PhysRevA.78.013614}, and include:
\begin{enumerate}
\item an operator-state correspondence, which relates the scaling dimensions of primary operators in free space to the energy levels (in units of the trap frequency) of the system confined to a harmonic trap;
\item virial theorems for trapped systems, which relate the expectation value of the potential energy operator to the expectation value of the untrapped Hamiltonian (as well as their powers);
\item and a tower of states in the trapped system which are separated by $2\omega$ (i.e., breathing modes).
\end{enumerate}
These properties are all generic, holding for both the few- and many-body systems, and may be explored and verified numerically via Monte Carlo simulations.

As will be discussed at greater length in \Sec{world-line_representation}, one of the main virtues of the one-dimensional system is that numerical simulations are unhampered by the algorithmic limitations that are often confronted in three dimensions.
Most notably, in contrast with the three-dimensional Fermi gas, numerical studies of the one-dimensional system can be performed completely free of sign and signal/noise  problems irrespective of any population- or mass-imbalance.
Given the striking qualitative similarities between the one- and three-dimensional unitary Fermi gases, one might hope to gain new qualitative insights about the latter (particularly within the physical regime that is currently inaccessible by numerical means) from quantitative numerical studies of former.
On a perhaps more speculative note, should a rigorous connection such as a duality between the theories eventually be established, one might envision that a prescription exists for relating one-dimensional observables (such as the Bertsch parameter and contact) to three-dimensional observables.
To that end, this work may be viewed as an initial step toward cataloging the properties of the one-dimensional system.

In this paper, I perform detailed numerical studies of few- and many-body systems confined to a harmonic trap and finite box, and present continuum limit extrapolated results for their energies and integrated contact densities.
In \Sec{theory}, I summarize the salient features of the lattice description used for this study, discuss the role of parameters appearing in the lattice theory, parameter tuning and the continuum limit.
In \Sec{world-line_representation}, I introduce a world-line representation for the partition function, and present explicit definitions for the physical observables of interest, including the energy of the system and integrated contact density (contact).
In \Sec{simulation_details}, I present details pertaining to the algorithm used to simulate the theory, the simulation parameters considered, and the generation of statistical ensembles.
In \Sec{analysis_and_results}, I discuss the analysis of data, including details relating to how continuum, infinite volume, and thermodynamic limit extrapolations were performed.
In this section I present continuum extrapolated estimates for few- and many-body observables confined to a trap and a box, and for the case of  many-body systems, also present thermodynamic limit estimates for the energies and integrated contact densities.

The many-body results for the Bertsch parameter presented in \Sec{analysis_and_results} were originally reported in \cite{PhysRevLett.109.250403}.
In this paper, I go into greater detail regarding the analysis of those results, as well as  present new results for the Bertsch parameter and integrated contact densities using ensembles of increased size.
From the latter estimates, I determine the universal parameter $\zeta$ as well as make a third determination of the Bertsch parameter using additional input from a density-functional theory calculation.
In \Sec{conclusion}, I summarize the results of this study and provide some concluding remarks.
Finally, in \App{thomas-fermi}, I derive the dependence of the trapped many-body energy on the parameters $\xi$ and $\zeta$, which were introduced in \Eq{bertsch_function}, and provide a confirmation of \Eq{trapped_energy} in the unitary limit.

\section{Theory}
\label{sec:theory}

The starting point for this study is an effective field theory for nonrelativistic fermions interacting via an attractive four-body contact interaction.
The continuum Lagrangian for the theory, defined in two-dimensional Euclidean space-time with temporal extent $\beta$ and spatial extent $L$, is given by:
\begin{eqnarray}
\calL = \psi^\dagger \left( \partial_\tau - \frac{\nabla^2}{2m} + v \right) \psi - \frac{g}{4!} (\psi^\dagger \psi)^4 \ ,
\label{eq:continuum_lagrangian}
\end{eqnarray}
where $\psi_\sigma(\tau,x)$ is a four-component Grassmann-valued spinor with spin components labeled by the index $\sigma=(a,b,c,d)$ and space-time coordinates labeled by the coordinate pair $(\tau,x)$, $m$ is the fermion mass, and $g$ is a coupling associated with the four-body interaction.
In addition, a spin-independent external potential $v(x)$ is introduced, and given by the two choices:
\begin{eqnarray}
v(x) = \left\{
\begin{array}{ll}
0 & \qquad \textrm{(untrapped)} \\
\frac{\kappa}{2} x^2 & \qquad \textrm{(trapped)} \ ,
\end{array}
\right. 
\label{eq:potential}
\end{eqnarray}
where the parameter $\kappa$ denotes the oscillator spring constant associated with a trapping potential.
I consider the theory at a finite temperature $T=\beta^{-1}$, employing anti-periodic boundary conditions in the time direction, and consider a system with open boundary conditions in the space direction.
The boundary condition choice for the latter is arbitrary, with different choices leading to different finite volume artifacts which are ultimately removed in the infinite volume limit.

The continuum theory is discretized on an $N_\tau \times (2 N_s+1)$ rectangular lattice with lattice sites labeled by the integer coordinate pair $\bfn = (n_\tau, n_s)$ for $n_\tau \in [0,N_\tau)$ and $n_s \in [-N_s,N_s]$.
In lieu of continuous fields, one considers fermion fields $\psi_\bfn$ (and their Hermitian conjugates $\psi_\bfn^\dagger$) defined only at the sites of the lattice.
The continuum operators appearing in \Eq{continuum_lagrangian} are then defined on the lattice using conventional finite difference discretizations, following \cite{PhysRevLett.92.257002}:
\begin{eqnarray}
\partial_\tau \psi + v\psi  & \to  & \frac{1}{b_\tau} \left( \psi_\bfn - e^{-b_\tau v_\bfn} \psi_{\bfn-{\bfe_\tau}} \right) \ ,\cr
-\nabla^2 \psi               & \to  & \frac{1}{b_s^2} \left( 2\psi_\bfn -\psi_{\bfn+{\bfe_s}} - \psi_{\bfn-{\bfe_s}} \right) \ ,\cr
\psi^\dagger \psi            & \to  & \psi_\bfn^\dagger \psi_{\bfn-{ \bfe_\tau}} \ ,
\label{eq:discretization}
\end{eqnarray}
where $b_\tau$ ($b_s$) is the temporal (spatial) lattice spacing with $\tau \equiv b_\tau n_\tau$ ($x \equiv b_s n_s$), and $\bfe_\tau$ ($\bfe_s$) is a unit vector pointing in the time (space) direction.
The lattice-discretized external potential is given by $v_\bfn = \frac{\kappa}{2} (b_s n_s)^2$.
The physical spatial extent of the lattice is given by $L = b_s (2 N_s+1)$, and the temporal extent (i.e., inverse temperature) is given by $\beta = b_\tau N_\tau$.

At infinite volume and at zero temperature, the ground state energy $E$ of the untrapped four-body system can be analytically related to the four-body coupling by exact diagonalization of the four-body transfer-matrix.
On the lattice and for positive couplings, the ground-state energy is given by solutions to the integral equation:
\begin{eqnarray}
\frac{1}{2\pi \hat g} =  \int_{-\pi}^\pi \left( \prod_\sigma{\frac{d \hat p_\sigma}{2\pi}} \right)  \frac{\delta(\sum_\sigma \hat p_\sigma) }{e^{-\hat E} \prod_\sigma \xi_{\hat p_\sigma}(\hat m) -1}\ ,
\label{eq:four_body}
\end{eqnarray}
where $\xi_{\hat p}(\hat m) = 1 + \Delta_{\hat p}/\hat m$, and $\Delta_{\hat p} = 2 \sin^2({\hat p}/2)$, $\hat g = b_\tau g/b_s^3$, $\hat E = b_\tau E$, $\hat p = b_s p$ and $\hat m = m b_s^2/b_\tau$\footnote{Throughout this work, I designate dimensionful quantities measured in lattice units with a caret.}.
One may define a four-particle scattering length $a$ by evaluating the scattering amplitude $\calA(p)$ for four-particle scattering at vanishing external momentum $p$, and requiring:
\begin{eqnarray}
\calA^{-1}(0) = \frac{m}{4\pi a} \ .
\end{eqnarray}
The four-body coupling may then be related to the scattering length by explicit evaluation of the inverse scattering amplitude on the lattice.
Doing so yields the relation:
\begin{eqnarray}
-\frac{\hat m}{4\pi \hat a} = \frac{1}{\hat g} - \frac{1}{\hat g_c}\ ,
\label{eq:coupling_scattering_length}
\end{eqnarray}
where $\hat a = a/b_s$, and $\hat g_c$ is obtained by evaluating \Eq{four_body} at vanishing binding energy.
Note that for an attractive coupling, by combining \Eq{four_body} and \Eq{coupling_scattering_length}, one obtains to leading order in $1/a$ (after restoring the lattice spacings) a four particle binding energy:
\begin{eqnarray}
-E = \frac{1}{2 m a^2} + \ldots\ .
\label{eq:binding_energy}
\end{eqnarray}
This result is very much analogous to that of two particles in three dimensions at large positive scattering length, and up to a constant of proportionality follows simply from dimensional analysis.
The unitary limit corresponds to tuning the scattering length to infinity, or correspondingly, the coupling $g$ to some $\calO(1)$ critical value $g_c$.

\begin{figure}
\includegraphics[width=0.4\textwidth]{\figdir 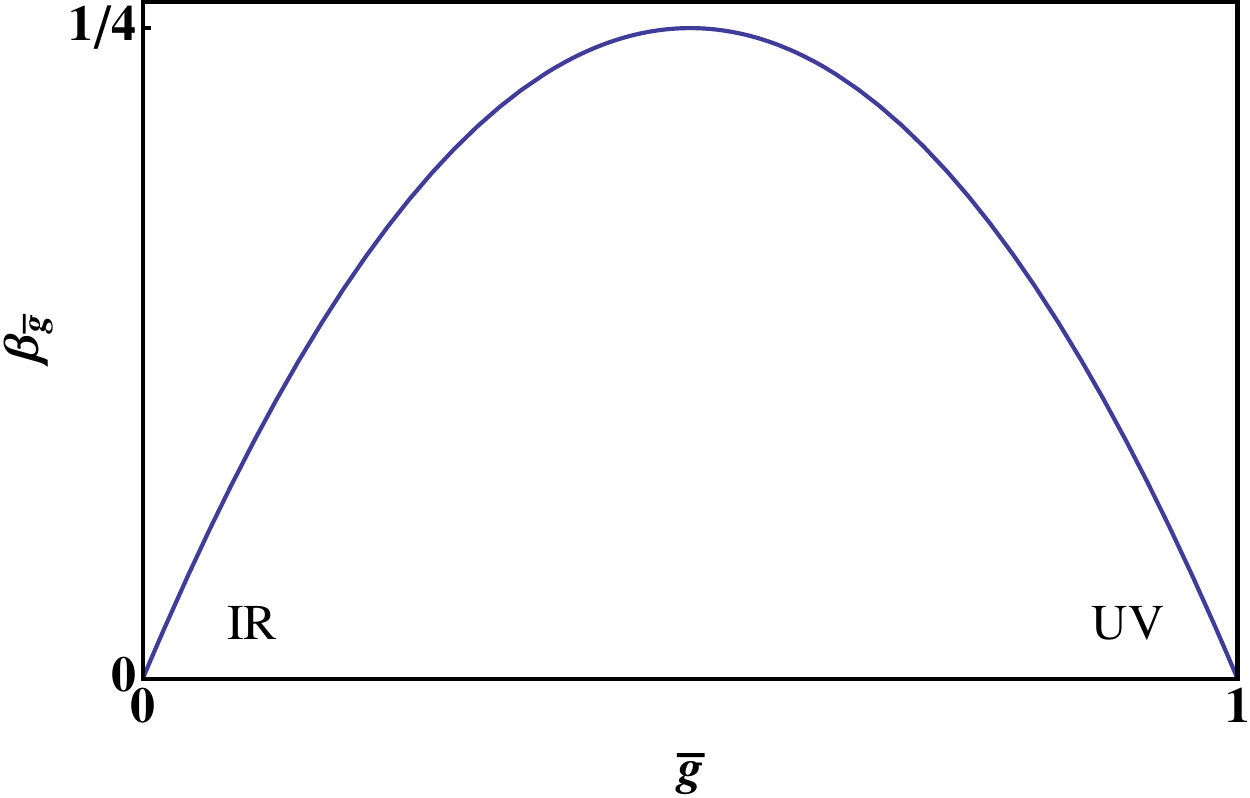}
\caption{
Beta function, $\beta_{\bar g}$, plotted as a function of the rescaled bare-coupling $\bar g = g/g_c$.
}
\label{fig:betafunc}%
\end{figure}

Note that the physical mass $m$ serves as a conversion factor between units of length and time, and so one may take $b_\tau \propto b_s^2$ providing $\hat m$ is held fixed.
The beta function
\begin{eqnarray}
\beta_{\bar g} = -\frac{d \bar g}{d \log b_s}\ ,
\label{eq:betafunc}
\end{eqnarray}
can then be computed for the rescaled bare coupling $\bar g \equiv g/g_c$ using \Eq{coupling_scattering_length}, and by requiring that the physical mass and scattering length be invariant under changes of the lattice spacing \cite{1996NuPhB.478..629K}.
The result is given by $\beta_{\bar g} = -\bar g (\bar g-1)$ and is plotted in \Fig{betafunc}.
A continuum theory may be defined at the zeros of the beta function;
in this case, one finds that there are two fixed points: a trivial one at vanishing coupling in the infra-red (IR) corresponding to the free theory, and a nontrivial fixed point at $g = g_c$ in the ultra-violet (UV), corresponding to the unitary limit.
The system is conformal and scale-invariant at both fixed points, and as such, no physical scales are available to characterize the system in those limits.

For this study, I am primarily interested in the nontrivial fixed point located at $g=g_c$.
Working in the canonical ensemble, every system of fixed total fermion number $Q$ is expected to have a zero-energy ground state and a vanishing integrated contact density.
This result simply follows from the fact that there are no scales in the problem, and therefore all dimensionful quantities must vanish.
Throughout this work, however, I consider systems of fixed fermion number confined to either harmonic trap or a finite box.
In such cases, scale invariance is explicitly broken by a new length scale that enters into the problem, namely, the characteristic size of the system.
To unify the discussion for trapped and untrapped fermions, I define the characteristic size of the system by $L_0 = (m \kappa)^{-1/4}$ (trapped) and $L_0 = 4L/\pi$ (untrapped).
From dimensional analysis considerations, the energy of the system must be proportional to the characteristic energy scale $\omega = 1/ (m L_0^2)$, and the integrated contact density must be proportional $1/L_0$.
This is true for both the free and unitary Fermi gas, although, the constants of proportionality will generally differ in each case.
Only in the former are the proportionality constants exactly calculable, and in the latter case they must be determined nonperturbatively.
As discussed earlier, in the unitary limit, the energy of the many-body trapped and untrapped systems are given by \Eq{untrapped_energy} and \Eq{trapped_energy}, respectively, where the constants of proportionality involving $\xi$ are undetermined.
Using the above definitions for $L_0$, the free-gas energies are given by \Eq{free_trapped_gas_energy} (trapped) and $E_0(Q) = Q^3 \omega /6$ (untrapped).
At finite volume and in the free theory limit, the integrated contact density vanishes both for trapped and untrapped systems.
In the latter case for the many-body system, this may be seen by simply differentiating \Eq{bertsch_function} with respect to the inverse scattering length and then taking the $a\to 0^-$ limit.

In the lattice theory, additional scales appear besides $L_0$ which also violate the scale-invariance of the continuum theory, namely the temporal and spatial lattice spacings.
One may quantify the lattice discretization errors using the dimensionless parameters $\epsilon_s = b_s/L_0$ and $\epsilon_\tau = b_\tau \omega \equiv \hat m \epsilon_s^2$.
From the latter it is evident that one may independently take the temporal continuum limit while holding the spatial lattice spacing fixed by considering the limit $\hat m \to \infty$.
For this study, I fix the anisotropy of the lattice (i.e., fixing $\hat m$ throughout the study) and then extrapolate the characteristic system size (in lattice units) to infinity, or equivalently $\epsilon_s\to0$.
This procedure allows one to take both spatial and temporal continuum limits (and infinite volume limits) simultaneously.
Further discussion of the lattice discretization errors and their removal may be found in \Sec{analysis_and_results}.

\section{World-line representation}
\label{sec:world-line_representation}

The partition function for the lattice theory is defined as a path-integral over fermion fields, weighted by the exponential of the lattice action (often with a chemical potential introduced to bias the fermion species numbers toward a desired value).
Conventional approaches for numerical simulation of the partition function require first reducing the action to a fermion bilinear via a Hubbard-Stratonovich transformation (i.e., the introduction of non-dynamical auxiliary fields) \cite{1957SPhD....2..416S,PhysRevLett.3.77}.
In the case of the four-body interaction appearing in the lattice expression for \Eq{continuum_lagrangian}, this can most easily be achieved using a discrete $Z_4$ field coupled to $\psi^\dagger_\bfn \psi_{\bfn-\bfe_\tau}$ (although there are other equally valid methods as well).
One then ``integrates out'' the fermion degrees of freedom leaving a path-integral over bosonic auxiliary degrees of freedom weighted by the exponential of a nonlocal action involving the logarithm of a fermion determinant.
The resulting effective action is generically complex, rendering standard importance sampling techniques which require a probabilistic interpretation for the path-integral measure inapplicable.
Phase reweighting and other techniques, while in principle may be applied to circumvent the problem, are in most cases prohibitively costly from the standpoint of computational resources and time due to signal/noise, and other problems.

It was recently demonstrated that a nonrelativistic four-component Fermi gas in one spatial dimension could be simulated on a lattice free of sign problems by considering alternative representations for the partition function \cite{2012PhRvA..85f3624E}.
For this study, I use a path-integral representation which was inspired by the so-called hopping parameter expansion \cite{itzykson:book}.
In this approach, one may express the partition function for the lattice theory as a path-integral over all possible self-avoiding time-directed fermion world-lines.
The representation is free of sign problems irrespective of population and mass imbalances, making it ideally suited for numerical study of few- and many-body four-component fermion systems.
Here, I briefly summarize the salient features of this formulation; for a more in-depth discussion, see \cite{2012PhRvA..85f3624E}.

\begin{figure}
\includegraphics[width=0.4\textwidth]{\figdir 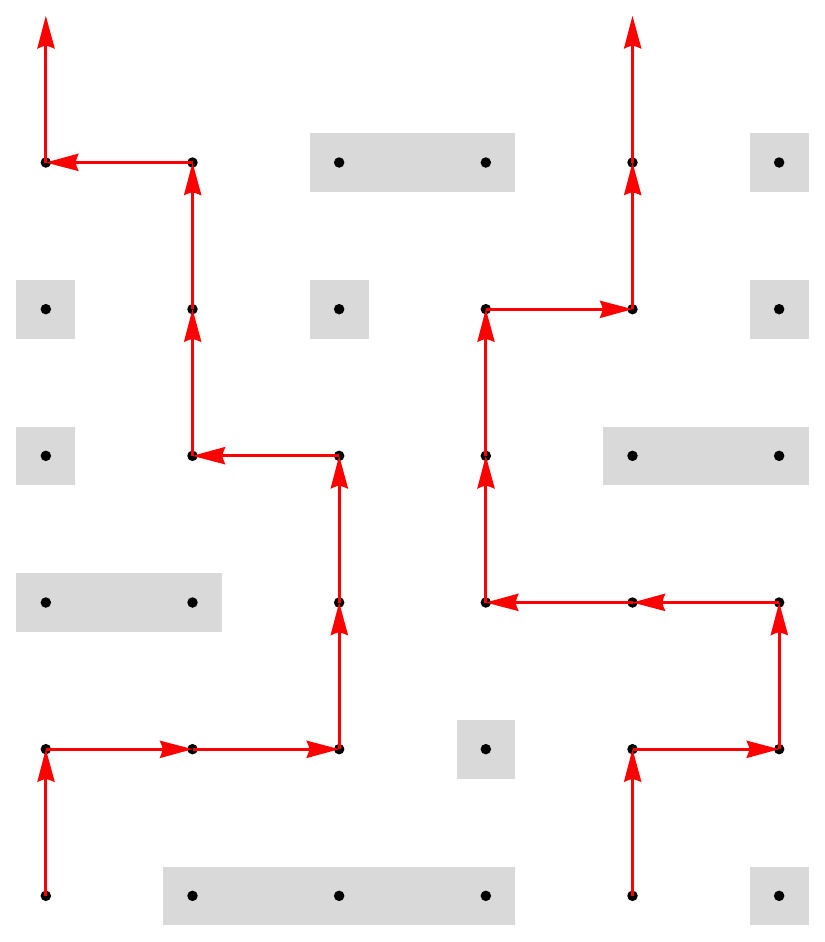}
\caption{A possible configuration $c\in C^*(2)$ for an $N_\tau = 2N_s+1 = 6$ lattice, reproduced from \cite{2012PhRvA..85f3624E}.
The set of 11 shaded domains $d$ represent $\calD(c)$; seven of those domains have $\calL(d) = 1$, three have  $\calL(d) = 2$ and one has  $\calL(d) = 3$. }
\label{fig:config}%
\end{figure}

For this study, I consider the canonical partition function $Z(q) \equiv e^{-\beta F(q)}$ for a four-component system comprising a fixed number of fermions $q_\sigma$ for each species $\sigma$, and having total fermion number given by $Q = \sum_\sigma q_a$.
In the world-line representation, the canonical partition function is given by the path-integral:
\begin{eqnarray}
Z(q) = \sum_{ c_\sigma \in C^*(q_\sigma) } \left[ \prod_\sigma  \left( \prod_{d\in\calD(c_\sigma)} z_{\calL(d)}(\hat m) \right) \left(\frac{1}{2 \hat m} \right)^{\calB_s(c_\sigma)}  e^{- N_\tau \calV(c_\sigma) } \right] \left(1 + \hat g \right)^{\calB_\tau( \cap_\sigma c_\sigma) } \ ,
\label{eq:partition_function}
\end{eqnarray}
where
\begin{eqnarray}
z_n(\hat m) = \frac{1}{2^{n+1}\sqrt{1+2/\hat m}} \left[ \left(1 + \frac{1}{\hat m} + \sqrt{1+\frac{2}{\hat m}}  \right)^{n+1} -  \left(1 + \frac{1}{\hat m} - \sqrt{1+\frac{2}{\hat m}}  \right)^{n+1}  \right]\ ,
\end{eqnarray}
and $C^*(q)$ is the set of all possible self-avoiding loops directed forward in time with a fixed winding number $q$.
An example of such a configuration for a single species is provided in \Fig{config}.
For a given configuration $c\in C^*(q)$, $\calB_\tau(c)$ represents the total number of time-like links associated with the configuration,\footnote{Note that in general $\calB_\tau(c) = N_\tau q$ for every $c\in C^*(q)$.} and $\calB_s(c)$ represents the number of space-like links associated with the configuration.
$\calD(c)$ represents the set of all maximal space-like linear domains $d$ of length $\calL(d)$ formed from the unoccupied sites of $c$; in \Fig{config}, such domains are indicated by the shaded gray regions.
Finally, the potential term in \Eq{partition_function} is given by
\begin{eqnarray}
\calV(c) = \frac{1}{N_\tau} \sum_{\tilde\bfn \in c} \hat v_{\tilde\bfn-\frac{1}{2}\bfe_\tau}\ ,\qquad \tilde\bfn = \bfn+\frac{1}{2}\bfe_\tau\ ,
\label{eq:potential_observable}
\end{eqnarray}
where $\hat v_\bfn = b_\tau v_\bfn$.

In this study, I focus on two physical quantities as a function of the fermion population: the energy of the system, $E(q) \equiv \lim_{\beta\to\infty} F(q)$, where $F(q)$ is the free-energy of the system, and the (integrated) contact density discussed in \Sec{introduction}.
Although Monte Carlo simulations provide a powerful tool for estimating observable quantities, the free-energy of the system itself is generally inaccessible due to the nature of algorithms employed.
As previously noted, however, for fermions in the unitary regime, the only physical length scales in the problem are the scattering length $a$ and box size $L_0$, up to discretization errors.
Furthermore, the mass parameter is the only quantity available for converting length scales into energy scales.
One may exploit these observations to gain access to the energy by noting that ground-state energy of the system must be proportional to the inverse fermion mass, implying
\begin{eqnarray}
E(q) = \lim_{\beta\to \infty} \frac{d F(q)}{d\log m} \ ,
\end{eqnarray}
up to finite lattice discretization errors.

The energy of the system may be determined using the Feynman-Hellmann theorem, which may be  expressed in the path-integral language as:
\begin{eqnarray}
\frac{d F(q)}{d \log m} = \frac{1}{\beta} \left\langle \frac{d S}{d \log m} \right\rangle_q\ ,
\label{eq:feynman-hellmann}
\end{eqnarray}
where $\langle\ldots\rangle_q$ is an expectation value taken with respect to a fixed-charge ensemble associated with $Z(q)$.
Note that the full derivative with respect to $\log{m}$ can be written in terms of partial derivatives with physical length scales held fixed.
Using the relation
\begin{eqnarray}
\frac{d}{d \log m} = \frac{\partial}{\partial \log m} + \left( \frac{\partial \log \kappa}{\partial \log m} \right)_{L_0} \frac{\partial}{\partial \log \kappa} + \left( \frac{\partial \log g}{\partial \log m}\right)_{a} \frac{\partial}{\partial \log g} \ ,
\end{eqnarray}
to differentiate the action in \Eq{feynman-hellmann}, one obtains three contributions to the energy:
\begin{eqnarray}
E(q)  = T(q) + V(q) + I(q)\ ,
\label{eq:energy}
\end{eqnarray}
in the zero temperature limit.
These contributions may be identified as the kinetic ($T$), potential ($V$) and interaction ($I$) energies, and correspond to partial differentiation of the the action with respect to $\log(1/m)$, $\log \kappa$ and $\log g$, respectively.
In the fermion world-line representation, the energy operators are explicitly given by:
\begin{eqnarray}
b_\tau T(q) = \lim_{N_\tau\to\infty} \frac{1}{N_\tau} \sum_\sigma   \left\langle \sum_{d\in \calD(c_\sigma)} \frac{\partial   }{\partial\log \hat m} \log \frac{ z_{\calL(d)}(\hat m) }{ z_{2N_s+1}(\hat m) }  -  \calB_s(c_\sigma)  \right\rangle_q \ ,
\end{eqnarray}
\begin{eqnarray}
b_\tau V(q) = \lim_{N_\tau\to\infty} \frac{1}{N_\tau} \sum_\sigma   \left\langle \calV(c_\sigma) \right\rangle_q \ ,
\end{eqnarray}
and
\begin{eqnarray}
I(q) = \left(- \frac{\partial \log \hat g}{\partial \log \hat m} \right) I_0(q)\ ,\qquad b_\tau I_0(q) =  -\frac{\hat g}{1+\hat g} \lim_{N_\tau\to\infty}  \frac{1}{N_\tau}    \left\langle \calB_\tau( \cap_\sigma c_\sigma) \right\rangle_q \ .
\label{eq:interaction_observable}
\end{eqnarray}
An expression for the prefactor appearing in \Eq{interaction_observable} for the interaction energy operator $I(q)$ at finite scattering length may be derived explicitly by differentiating both sides of \Eq{coupling_scattering_length} with respect to $\hat m$ while holding all physical length scales fixed.
Doing so yields the useful relation: 
\begin{eqnarray}
\frac{1}{\hat g} \left(  \frac{\partial \log \hat g}{\partial \log \hat m} +1 \right)  = \frac{1}{\hat g_c} \left(  \frac{\partial \log \hat g_c}{\partial \log \hat m} + 1 \right)\ .
\end{eqnarray}
Taking the temporal continuum limit with $b_s$ held fixed (i.e., $\hat m \to \infty$), one finds that $\partial \log \hat g / \partial \log \hat m \to -1$, and therefore $I(q) \to I_0(q)$.
For any finite anisotropy, however, the prefactor remains nontrivial and its inclusion is crucial for obtaining correct continuum limit estimates.

Starting from \Eq{contact_definition} and using \Eq{coupling_scattering_length}, the integrated contact density $C(q)$ (not to be confused with the set of closed-loop configurations $C^*$ discussed above) may be written as
\begin{eqnarray}
C(q) = -(m g)^2 \frac{d E(q)}{d g}\ .
\label{eq:contact}
\end{eqnarray}
Taking $E(q)$ as the zero temperature limit of the logarithm of the canonical partition function, I obtain
\begin{eqnarray}
C(q) = -\frac{1}{g} (mg)^2  I_0(q)
\label{eq:contact1}
\end{eqnarray}
for the contact.
One can derive other expressions for the contact by combining \Eq{contact} and, for example, \Eq{energy} (as opposed to the logarithm of the partition function).
Such expressions are expected to yield the same continuum limit as \Eq{contact1}, although estimates based upon such formulas are presumably noisier since they rely on the correlations among the various energy observables.
For this reason, I only consider estimates of the contact based on \Eq{contact1}.

\section{Simulation details}
\label{sec:simulation_details}

\begin{figure}
\includegraphics[height=2.0cm]{\figdir 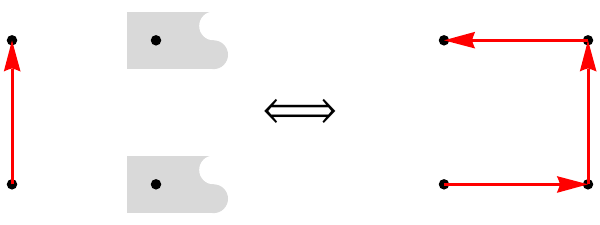} \hspace{36pt}
\includegraphics[height=2.0cm]{\figdir 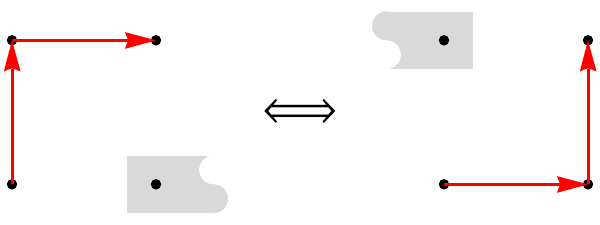} 
\caption{Two of the four allowed local constraint-preserving configuration updates (the remaining two possibilities are just mirror images of those shown above).}
\label{fig:updating}%
\end{figure}

Monte Carlo simulations were performed for fermions confined to a finite box and a harmonic trap.
Ensembles were generated using a local updating scheme which preserves the constraints placed on the configuration space.
Particularly, time-directed links were updated one at a time in accordance with the updating rules depicted in \Fig{updating}.
Proposed local updates were either accepted or rejected using a Metropolis accept/reject step.
Whenever a proposed update violated the constraints on allowable configurations (such as constraints imposed by Pauli exclusion, or the lattice boundaries) those proposed updates were rejected with unit probability.
It is known that local updating schemes generically suffer from critical slowing, and this updating scheme is by no means any different.
However, for the lattice volumes and physical parameters explored in this study, the efficiency of the updating scheme was found to be adequate for achieving percent-level estimates of observables with available computational resources.

All random numbers used in the simulations were generated using L\"{u}scher's Ranlux pseudo-random number generator with a luxury level equal to one \cite{Luscher1994100}.
For this study, given the simplicity of the updating scheme, it was found that the random number generation was the most time-consuming part of the simulations.
The random number generators used to generate the configurations for each ensemble were independently seeded so-as to yield uncorrelated ensembles.
Due to inefficiencies in the Monte Carlo algorithm, however, each ensemble involved configurations which were highly correlated in Monte Carlo time.
Generally, the autocorrelations in each ensemble depend strongly on the simulation parameters considered, and so care was taken to prune the ensembles so-as to eliminate such correlations.


\begin{table}
\caption{%
\label{tab:few-body_trapped_results}%
Continuum extrapolated observables for up to eight fermions confined to a harmonic trap.
}
\begin{ruledtabular}
\begin{tabular}{c|cccc|cc|cc|cc|cc}
$Q$ & $q_a$ & $q_b$ & $q_c$ & $q_d$ & $E/\omega$ (I) & $\chi^2$/d.o.f. & $E/\omega$ (II) & $\chi^2$/d.o.f. & $E/\omega$ (III) & $\chi^2$/d.o.f. & $C L_0$ & $\chi^2$/d.o.f. \\
\hline
4  & 1 & 1 & 1 & 1 &  1.008(5) & 1.0  &  1.01(1) & 1.1    &  1.01(1) & 1.0  &  7.13(4) & 0.6  \\
\hline                                                                                       
5  & 2 & 1 & 1 & 1 &  2.341(6) & 0.1  &  2.34(1) & 1.0    &  2.34(1) & 0.5  &  7.55(5) & 0.7  \\
\hline                                                                                       
6  & 3 & 1 & 1 & 1 &  4.535(6) & 3.6  &  4.50(1) & 1.6    &  4.57(1) & 3.0  &  8.98(4) & 2.6  \\
   & 2 & 2 & 1 & 1 &  3.610(9) & 0.4  &  3.60(2) & 2.8    &  3.61(2) & 1.4  &  8.25(3) & 2.4  \\
\hline                                                                                       
7  & 4 & 1 & 1 & 1 &  7.813(6) & 1.1  &  7.80(2) & 1.1    &  7.83(2) & 1.1  &  9.62(4) & 1.2  \\
   & 3 & 2 & 1 & 1 &  5.796(7) & 1.6  &  5.79(1) & 1.2    &  5.80(2) & 2.0  &  9.30(3) & 0.8  \\
   & 2 & 2 & 2 & 1 &  4.719(7) & 0.9  &  4.71(1) & 1.5    &  4.72(2) & 0.3  &  8.87(3) & 1.3  \\
\hline                                                                                       
8  & 5 & 1 & 1 & 1 & 12.060(6) & 0.9  & 12.05(2) & 0.1    & 12.07(2) & 0.1  & 10.69(2) & 1.2  \\
   & 4 & 2 & 1 & 1 &  9.043(5) & 0.6  &  9.03(1) & 0.8    &  9.05(1) & 0.6  & 10.15(2) & 1.4  \\
   & 3 & 3 & 1 & 1 &  7.964(5) & 0.6  &  7.92(1) & 0.6    &  8.01(1) & 0.6  & 10.34(3) & 1.0  \\
   & 3 & 2 & 2 & 1 &  6.958(8) & 0.9  &  6.96(2) & 0.7    &  6.96(2) & 1.5  &  9.66(3) & 2.0  \\
   & 2 & 2 & 2 & 2 &  4.570(8) & 2.7  &  4.55(2) & 6.1    &  4.59(2) & 1.3  & 19.38(4) & 4.4  \\
\end{tabular}
\end{ruledtabular}
\end{table}

\begin{figure}[b]
\includegraphics[width=\figwidth]{\figdir 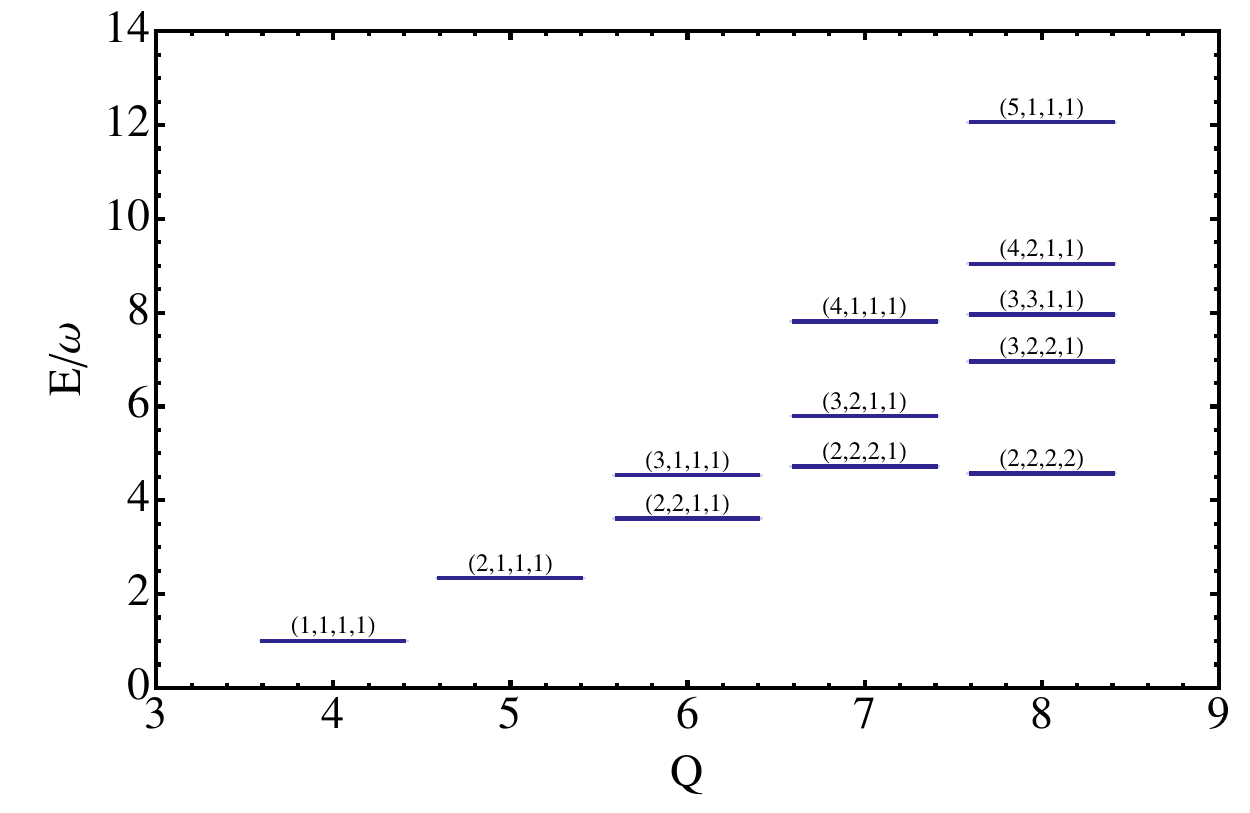} \,
\includegraphics[width=\figwidth]{\figdir 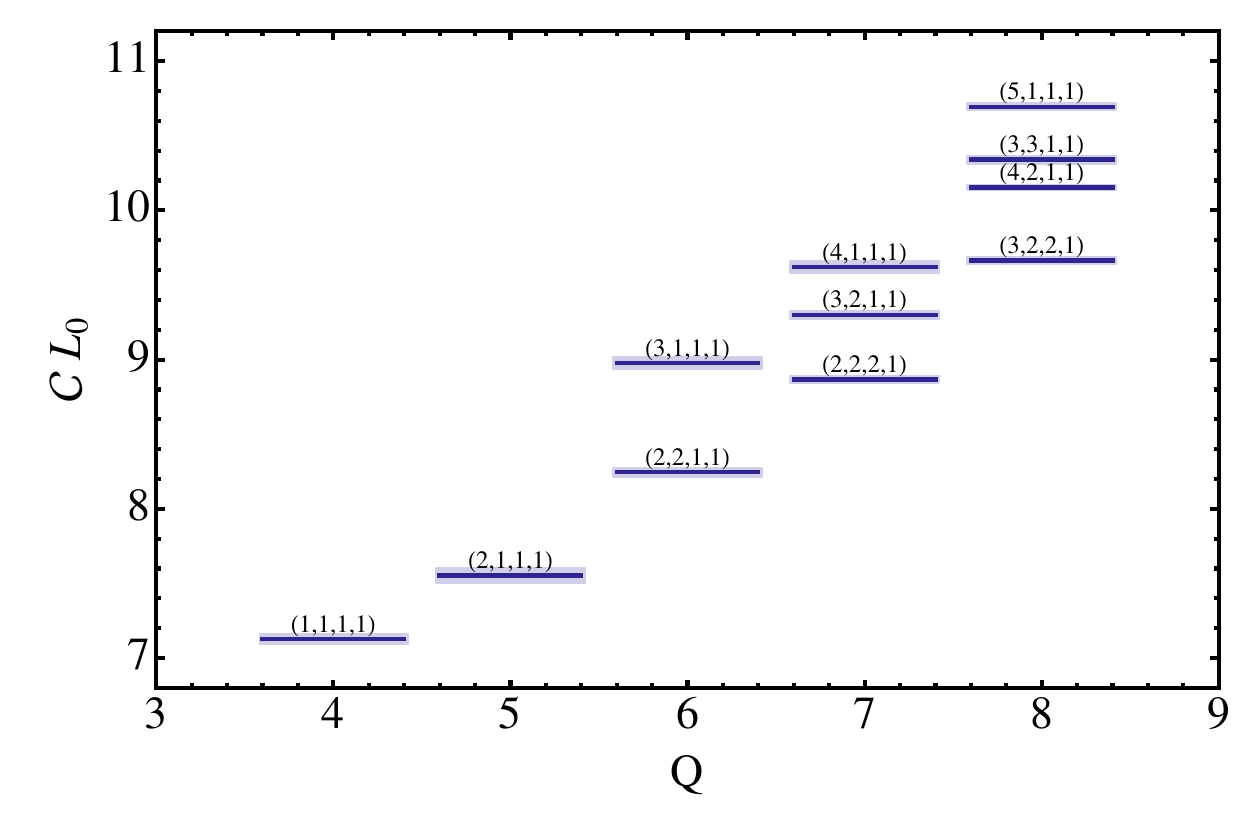}
\caption{%
Continuum few-body energies (obtained using definition I) and integrated contact densities for trapped fermions.
The result for the contact determined for $q=(2,2,2,2)$ has been omitted from the plot for clarity purposes, but can be found in \Tab{few-body_trapped_results}.
}
\label{fig:few-body_summary}%
\end{figure}

One interesting feature of the world-line path-integral representation presented in \Sec{world-line_representation} over conventional approaches that work with fermion determinants, is that in some situations there is no computational limitation on the spatial size of the lattice.
To see this, first note that the computational cost of updating a single configuration by sweeping through the lattice scales like $\beta Q$.
At low temperature, however, the computational cost of a single update scales implicitly like the square of whatever length scale in the problem is smallest.
This is because, roughly speaking, the smallest length scale is what determines the energy splittings in the system.
So for example, if the only length scale in the problem is the volume $L$, then in order to study the ground state properties of the system, one requires $\beta  \sim L^2$.
If smaller length scales are present, such as a characteristic trap size $L_0$ of a trapping potential or a finite scattering length $a$, one can then increase the spatial volume arbitrarily without increasing the computational cost of the simulation since the energy splittings are then determined by those other smaller scales.


\begin{table}
\caption{%
\label{tab:many-body_trapped_results}%
Continuum extrapolated observables for up to $56$ fermions confined to a harmonic trap.
}
\begin{ruledtabular}
\begin{tabular}{c|cccc|cc|cc|cc|cc}
$Q$ & $q_a$ & $q_b$ & $q_c$ & $q_d$ & $\xi_Q^{1/2}$ (I) & $\chi^2$/d.o.f. & $\xi_Q^{1/2}$ (II) & $\chi^2$/d.o.f. & $\xi_Q^{1/2}$ (III) & $\chi^2$/d.o.f. & $C L_0/Q^{3/2}$ & $\chi^2$/d.o.f. \\
\hline
12  & 3  &  3  &  3  &  3   & 0.592(3) & 0.8 & 0.589(3) & 1.4 & 0.591(6) & 0.5 & 0.856(4) & 1.0 \\
16  & 4  &  4  &  4  &  4   & 0.599(2) & 1.0 & 0.598(2) & 0.5 & 0.601(5) & 1.2 & 0.856(5) & 0.2 \\
20  & 5  &  5  &  5  &  5   & 0.605(2) & 1.3 & 0.598(2) & 0.8 & 0.607(4) & 0.8 & 0.863(4) & 0.4 \\
24  & 6  &  6  &  6  &  6   & 0.606(2) & 0.8 & 0.606(2) & 0.4 & 0.607(4) & 0.7 & 0.853(4) & 0.3 \\
28  & 7  &  7  &  7  &  7   & 0.610(2) & 0.9 & 0.608(3) & 1.4 & 0.611(4) & 0.4 & 0.869(3) & 1.0 \\
32  & 8  &  8  &  8  &  8   & 0.611(3) & 0.7 & 0.611(4) & 0.6 & 0.609(7) & 0.5 & 0.850(5) & 0.8 \\
36  & 9  &  9  &  9  &  9   & 0.612(2) & 0.9 & 0.608(3) & 0.1 & 0.616(4) & 0.8 & 0.856(7) & 0.9 \\
40  & 10 &  10 &  10 &  10  & 0.612(2) & 1.1 & 0.610(4) & 0.9 & 0.616(6) & 0.7 & 0.841(9) & 1.3 \\
44  & 11 &  11 &  11 &  11  & 0.613(1) & 0.7 & 0.613(3) & 0.6 & 0.612(4) & 0.7 & 0.866(6) & 0.8 \\
48  & 12 &  12 &  12 &  12  & 0.613(2) & 2.7 & 0.617(4) & 0.7 & 0.608(5) & 2.0 & 0.856(6) & 1.0 \\
52  & 13 &  13 &  13 &  13  & 0.612(2) & 0.7 & 0.617(3) & 0.3 & 0.607(4) & 0.9 & 0.869(5) & 0.3 \\
56  & 14 &  14 &  14 &  14  & 0.612(2) & 0.4 & 0.611(3) & 3.1 & 0.617(5) & 0.9 & 0.854(8) & 1.0 \\
\end{tabular}
\end{ruledtabular}
\end{table}

The trapped simulations for this study were performed on a finite lattice chosen such that $L\gg L_0$, where $L$ corresponds to the box size.
Generally finite volume errors for the trapped system depend on the likelihood for the few- or many-body ground state wavefunction to lie outside the box \cite{PhysRevA.84.043644}.
Given that the ground state  wavefunction for trapped unitary fermions behaves asymptotically like a harmonic oscillator wave function, one can expect finite volume artifacts to be exponentially suppressed in $L/L_0$.
All few- and many-body simulations for trapped fermions in this study were performed at spatial lattice volumes satisfying $L/L_0 \gtrsim 25$, and by the scaling arguments above may effectively be regarded as at infinite volume.
Although not done so in this study, one may easily monitor the configurations as they are updated and verify explicitly whether updates carry fermions to the edge of the box when a confining potential is present.
The probability for such an occurrence during the course of a simulation that has run for a finite amount of time is exponentially small in $L/L_0$.

Simulations of trapped and untrapped fermions were performed using temporal extents much larger than the expected inverse energy splittings of the system in order to ensure adequate suppression of excited state contamination (or thermal excitations).
For trapped few- and many-body systems, properties of the Schr\"{o}dinger algebra imply that the spectrum contains a tower of breathing modes, each separated by an amount $2\omega$ \cite{PhysRevD.76.086004}.
In the case of trapped fermions, the temporal extent of the lattice was therefore chosen to satisfy $\beta \omega \gtrsim 10$.
For untrapped many-body systems, the energy splittings are expected to be of order the Fermi energy, given by $E_F(Q) = Q^2 \omega/2$, and therefore the temporal extent of the lattice was chosen to satisfy $\beta E_F \gtrsim 10$.
Simulations were performed using a single fixed value of the lattice mass parameter, $\hat m = 1.3$, corresponding to a critical coupling $\hat g_c \approx 3.7237$

Simulations were performed for multiple values of $\epsilon_s$ in order to perform continuum limit extrapolations of observables estimated on ensembles of fixed fermion number.
Few-body trapped ensembles (i.e, $Q\lesssim 8$ for all possible $q$) were generated for all integer values of $1/\epsilon_s\in[3,12]$, and consisted of approximately $800$-$50000$ uncorrelated configurations after thermalization\footnote{In \cite{PhysRevLett.109.250403} I erroneously wrote that all simulations consisted of 150-350 configurations; this claim in fact only applied to the many-body simulations and not to the few-body cases $Q=4,5$.
The latter ensembles were essentially the same as those used in this study, and were considerably larger in size.}.
Note that fewer configurations were generated at smaller $\epsilon_s$ as a result of increased autocorrelation times due to critical slowing, and also because of the associated increase in $\beta$ with $L_0^2$.

\begin{figure}
\includegraphics[width=\figwidth]{\figdir 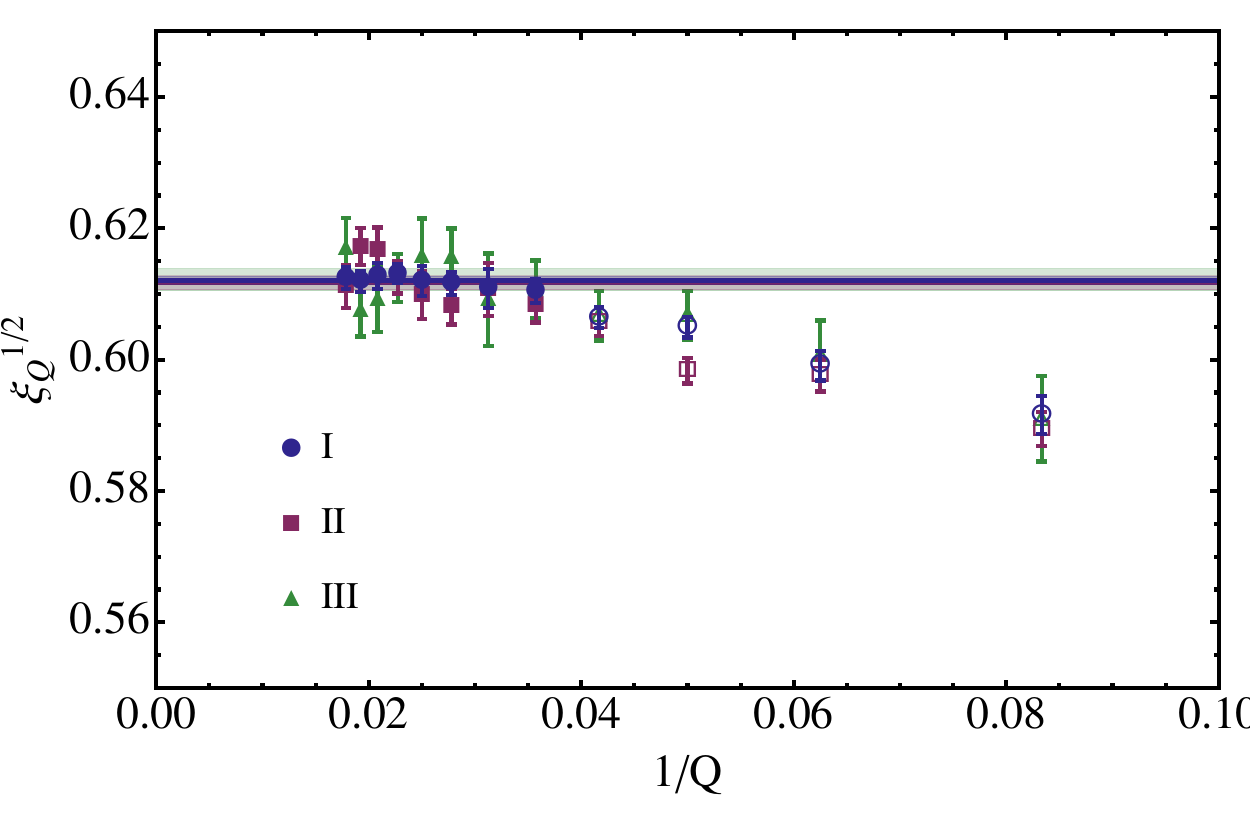} \,
\includegraphics[width=\figwidth]{\figdir 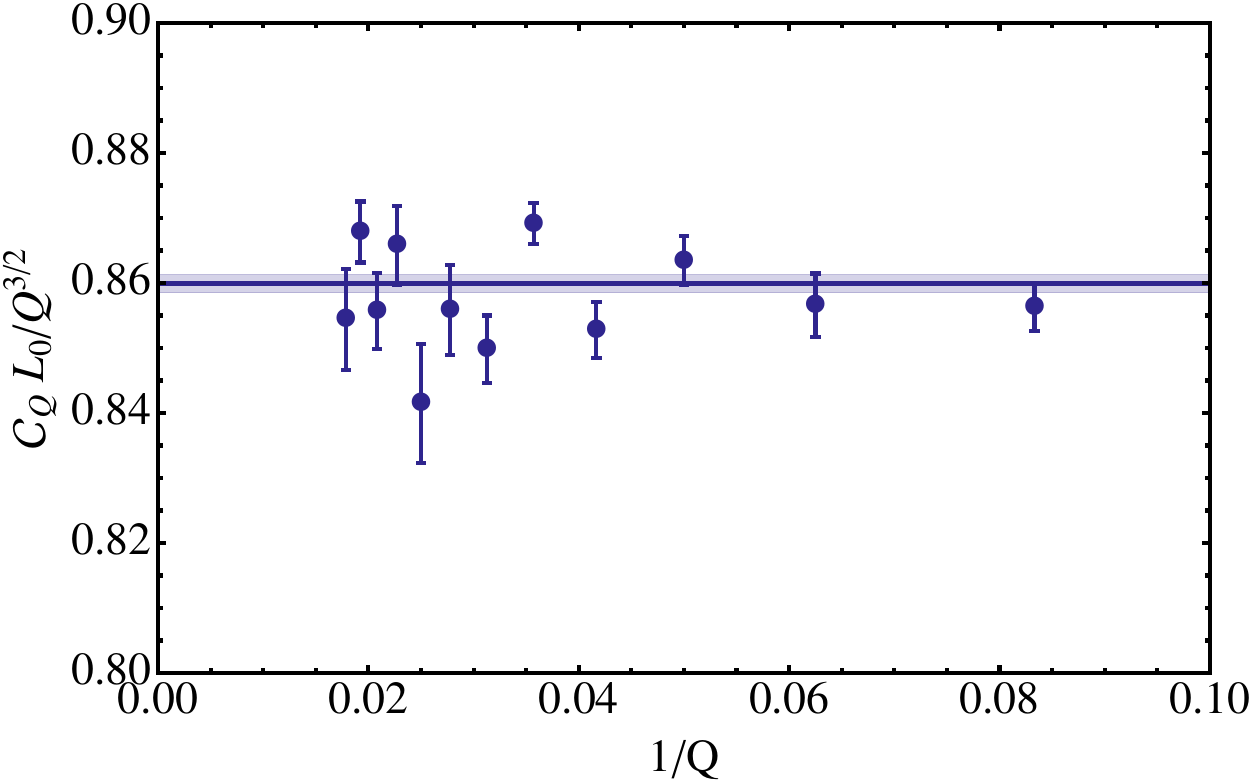}
\caption{
Thermodynamic limit extrapolation of continuum energies and integrated contact densities for the trapped Fermi gas.
}
\label{fig:trapped-thermo_extrap}%
\end{figure}

Trapped and untrapped many-body ensembles were the same as to those used in \cite{PhysRevLett.109.250403}, however, the size of the ensembles have been enlarged, particularly for the untrapped studies.
Trapped many-body ensembles were generated for $Q=28,32,36,40,44,48,52,56$ at integer values of $1/\epsilon_s \in[7,20]$, subject to the constraint $Q^{1/2}\epsilon_s \le 1.0$.
Ensembles consisted of approximately 200-1300 uncorrelated configurations with larger ensemble sizes corresponding larger $\epsilon_s$ and smaller $Q$.
Untrapped many-body ensembles were generated for $Q=32,48,56,64,72,80,88$ for equally spaced values of $k_F b_s = Q \epsilon_s \in[0.15,0.7]$.
Ensembles consisted of approximately 600-1600 uncorrelated configurations, again with larger ensemble sizes corresponding larger $\epsilon_s$ and smaller $Q$.

\section{Analysis and results}
\label{sec:analysis_and_results}

Finite lattice discretization errors may be understood from the view-point of a Symanzik action \cite{Symanzik1983187, Symanzik1983205}, a continuum description of the lattice theory with lattice spacing dependence carried by the undetermined couplings associated with higher dimension operators.
In principle, one should consider the inclusion of all possible local operators consistent with the underlying symmetries of the lattice action.
Since the continuum theory is scale-invariant, dimensional considerations imply that operators of scaling dimension $\Delta_{\calO}$ must have associated couplings with lattice spacing dependence that scales like $b_s^{\Delta_{\calO}-3}$, where again for a fixed physical mass and lattice mass parameter, I have used the fact that $b_\tau \sim b_s^2$.
Throughout this study, I consider dimensionful observables expressed in either units of the trap frequency or trap size.
Since quantum corrections to the continuum observables typically involve powers of coupling constants associated with higher-dimension operators, one may infer the $L_0$ dependence of such corrections by requiring that the final result be dimensionless.
Since $L_0$ is the only other dimensionful length scale in the problem besides the lattice spacings, one concludes that operators of scaling dimension $\Delta_\calO$ induce volume dependence scaling like $\epsilon_s^{\Delta_\calO-3}$ for dimensionless observables.

The scaling dimensions of the lowest-dimension few-body operators have been studied in detail in \cite{Nishida:2009pg} and confirmed numerically in \cite{PhysRevLett.109.250403}.
For the theory under investigation one concludes that dimensionless observables for a system of fixed total fermion number Q (such as the energy measured in units of $\omega$ or the contact in units of $L_0^{-1}$) must scale as
\begin{eqnarray}
\calO_Q(\epsilon_s) = \calO_Q + \calO^{(1)}_Q \epsilon_s + \calO_Q^{(5/3)} \epsilon_s^{5/3} + \ldots\ ,
\label{eq:fit_func-cont}
\end{eqnarray}
where $\calO_Q$ is the physical observable in the continuum limit (independent of $\hat m$), and $\calO_Q^{(j)}$ ($j = 1, 5/3, \ldots$) are unknown coefficients that depend implicitly on the dimensionless parameter $\hat m$.
The term linear in $\epsilon_s$ is the leading lattice spacing error attributed to an untuned $4\to4$ operator with a derivative insertion having scaling dimension $\Delta_\calO/2 = 2$.
The subleading correction is attributed to an parity-odd $5\to5$ operator with scaling dimension $\Delta_\calO/2 = 7/3$.
Note that the lattice action does not give rise to $2\to2$ or $3\to3$ interactions as a result of the point-split nature of the four-body interaction defined in \Eq{discretization}.

Continuum limit extrapolations of the dimensionless energies and integrated contact densities were performed for systems of fixed fermion number by fitting estimated observables to \Eq{fit_func-cont} truncated at order $\epsilon_s^{5/3}$.
For each observable, fits were performed over a range of $\epsilon_s$ values, and the maximum $\epsilon_s$ included in the fit was varied in order to evaluate the robustness of the extrapolation.
Plots of the fit results for all few- and many-body observables are provided in \cite{supp}.
Included are: 1) plots of the estimated observable $\calO_Q(\epsilon_s)$ for fixed $Q$ as a function of $\epsilon_s$, and the fit curve obtained using \Eq{fit_func-cont} and its associated error band, and 2) plots of the extrapolated fit value $\calO_Q$ and associated errors as a function of the maximum $\epsilon_s$ included in the fit.

For trapped systems, three different definitions of the energy were considered:
\begin{displaymath}
E = \left\{\begin{array}{ll}
T + V + I & \qquad  (I) \\
2 V             & \qquad (II) \\
2 ( T + I ) & \qquad (III) \\
\end{array}\right.
\end{displaymath}
Definitions (II) and (III) follow from the virial theorem for trapped unitary fermions, and definition (I), given by \Eq{energy}, is simply the average of the latter two.
Although the virial theorem is violated at finite $\epsilon_s$, in the continuum limit the three definitions are expected to converge.
One may either use the three definitions to confirm restoration of the virial theorem in the continuum limit for each fixed charge system, or one may use the three definitions to gauge the systematic errors in the extrapolations.
For trapped estimates of the energy, I indicate which energy definition is used by the Roman numerals (I), (II) and (III).
Energy estimates for untrapped systems use definition (I) with $V=0$, and all estimates of the contact use \Eq{contact1}.

\begin{table}
\caption{%
\label{tab:many-body_untrapped_results}%
Continuum extrapolated observables for up to $88$ fermions confined to a finite box.
}
\begin{ruledtabular}
\begin{tabular}{c|cccc|cc|cc}
$Q$ & $q_a$ & $q_b$ & $q_c$ & $q_d$ & $\xi_Q$ & $\chi^2$/d.o.f. & $C_Q L_0/Q^2$ & $\chi^2$/d.o.f. \\
\hline
32  & 8  &  8  &  8  &  8   & 0.432(2) & 0.4 & 1.225(6) & 1.8 \\
48  & 12 &  12 &  12 &  12  & 0.413(3) & 0.3 & 1.192(4) & 2.1 \\
56  & 14 &  14 &  14 &  14  & 0.406(2) & 0.9 & 1.183(3) & 0.4 \\
64  & 16 &  16 &  16 &  16  & 0.401(2) & 0.4 & 1.177(3) & 2.8 \\
72  & 18 &  18 &  18 &  18  & 0.397(2) & 1.1 & 1.162(5) & 1.0 \\
80  & 20 &  20 &  20 &  20  & 0.395(2) & 0.6 & 1.162(3) & 1.5 \\
88  & 22 &  22 &  22 &  22  & 0.393(2) & 1.0 & 1.154(6) & 0.7 \\
\end{tabular}
\end{ruledtabular}
\end{table}

\begin{figure}[b]
\includegraphics[width=\figwidth]{\figdir 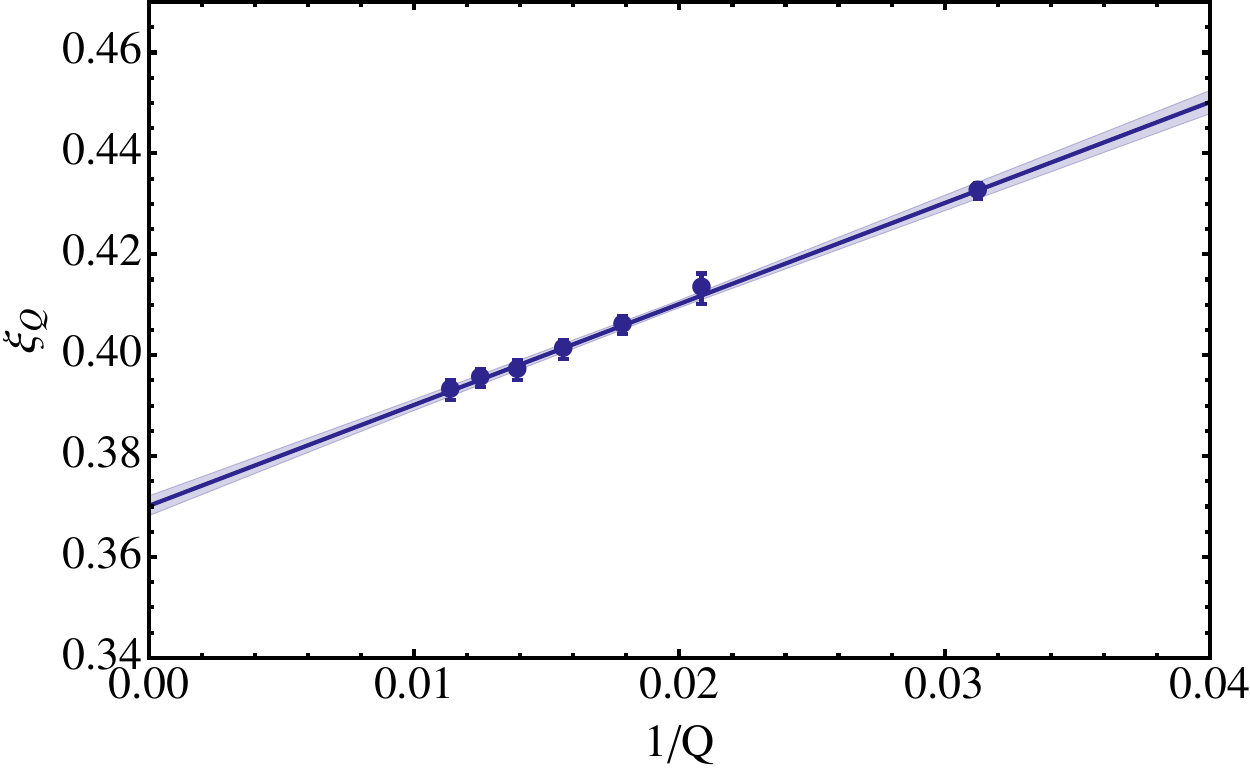} \,
\includegraphics[width=\figwidth]{\figdir 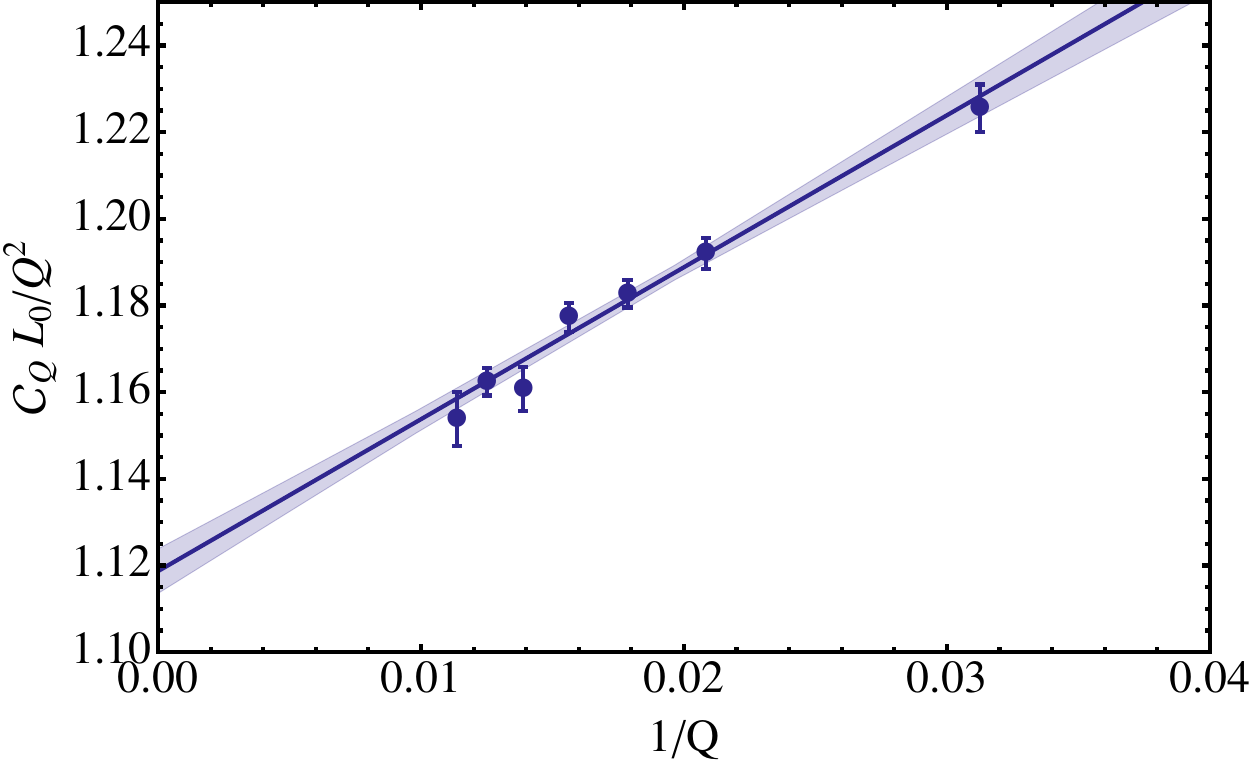}
\caption{
Thermodynamic limit extrapolation of continuum energies and integrated contact densities for the untrapped Fermi gas.
}
\label{fig:untrapped-thermo_extrap}%
\end{figure}

Continuum limit estimates for each trapped few-body observable are provided in \Tab{few-body_trapped_results} along with the corresponding $\chi^2$ per degree of freedom (d.o.f) as a measure of the goodness of fit.
Plots summarizing the fit results are also provided in \Fig{few-body_summary}.
For the two cases $Q=4,5$, the energies in units of $\omega$ are known analytically to be unity and $7/3$ respectively; these results follows from the operator-state correspondence and knowledge of the scaling dimensions of few-body operators \cite{Nishida:2009pg}.
Continuum extrapolations of the energies yield results statistically consistent with the exactly determined values to within 1\% and 0.5\% statistical errors, respectively. 
Extrapolation results for the Bertsch parameter, $\xi_Q$, and the integrated contact density, $C_Q$, defined at a finite $Q$, are tabulated in \Tab{many-body_trapped_results} and \Tab{many-body_untrapped_results} for the trapped and untrapped many-body systems.
Plots summarizing the continuum extrapolation results are shown in \Fig{trapped-thermo_extrap} and \Fig{untrapped-thermo_extrap}.

Thermodynamic limit extrapolations of the continuum observables $\xi_Q$ and $C_Q$ (appropriately normalized) were carried out for the many-body systems.
The leading dependence on $1/Q$ for these quantities is presently unknown, but is expected to be of the form
\begin{eqnarray}
\calO_Q = \calO + \calO^{(p)} Q^{-p} + \ldots\ .
\label{eq:fit_func-therm}
\end{eqnarray}
Following the approach of \cite{PhysRevLett.109.250403}, I use an ansatz fit function for the Bertsch parameter extrapolation.
For trapped fermions, I fix $\calO^{(p)} = 0$ and determine the parameter $\calO$ using a constant linear least-squares fit over a fit range in which observables appear independent of $Q$ (within statistical uncertainties).
For untrapped fermions, I use the ansatz $p=1$ and determine the fit parameters $\calO$ and $\calO^{(p)}$.
Fit results for each case are presented in \Tab{many-body_trapped_extrap} (trapped) and \Tab{many-body_untrapped_extrap} (untrapped) along with the fit range used, and the goodness of fit.
Fit results and associated error bands are plotted in \Fig{trapped-thermo_extrap} and \Fig{untrapped-thermo_extrap}.
Within this analysis, I obtain the Bertsch parameters $\xi=0.375(1)$ (I), $\xi=0.374(1)$ (II) and $\xi=0.375(2)$ (III) for trapped fermions,  and $\xi=0.370(2)$ for untrapped fermions.
These results are consistent to within about two standard deviations, and are also consistent with the analysis of \cite{PhysRevLett.109.250403}.
Similar fits were performed for the contact and yield the values $\lim_{Q\to\infty} C_Q L_0/Q^{3/2} = 0.860(1)$ (trapped) and $\lim_{Q\to\infty} C_Q L_0/Q^2 = 1.119(5)$ (untrapped) in the thermodynamic limit.

\begin{table}
\caption{%
\label{tab:many-body_trapped_extrap}%
Thermodynamic limit extrapolated observables for fermions confined to a harmonic trap.
Parameters without uncertainties may be regarded as fixed during the fitting procedure.
}
\begin{ruledtabular}
\begin{tabular}{l|ccccc}

                      & $\calO$ & $\calO^{(p)}$ & $p$ & $\chi^2$/d.o.f. & fit interval \\ 
\hline
$\xi_Q^{1/2}$ (I)          & 0.6121(6) & 0       & n/a     & 0.3     & 28-56 \\
$\xi_Q^{1/2}$ (II)          & 0.612(1) & 0       & n/a     & 1.6     & 28-56 \\
$\xi_Q^{1/2}$ (III)          & 0.612(2) & 0       & n/a     & 0.7     & 28-56 \\
$C_Q L_0 /Q^{3/2}$      & 0.860(1) & 0       & n/a     & 2.6     & 12-56 \\
\end{tabular}
\end{ruledtabular}
\end{table}

\begin{table}
\caption{%
\label{tab:many-body_untrapped_extrap}%
Thermodynamic limit extrapolated observables for fermions confined to a finite box.
Parameters without uncertainties may be regarded as fixed during the fitting procedure.
}
\begin{ruledtabular}
\begin{tabular}{l|ccccc}

                   & $\calO$ & $\calO^{(p)}$ & $p$ & $\chi^2$/d.o.f. & fit interval \\ 
\hline
$\xi_Q$             & 0.370(2) & 2.0(1)    & 1       & 0.1     & 32-88 \\
$C_Q L_0 /Q^2$      & 1.119(5) & 3.5(3)    & 1       & 1.0     & 32-88 \\
\end{tabular}
\end{ruledtabular}
\end{table}

Having determined the energies and integrated contact densities for the trapped and untrapped many-body systems, it is then possible to determine the subleading parameter $\zeta$ appearing in \Eq{bertsch_function}.
In the untrapped case the relationship is trivially given by
\begin{eqnarray}
\lim_{Q\to\infty} \frac{C_Q L_0}{Q^2} = \frac{2\pi}{3} \zeta\ .
\label{eq:trapped_zeta}
\end{eqnarray}
Plugging in the value for the contact obtained from \Tab{many-body_untrapped_extrap} yields $\zeta = 0.534(3)$.
A far less trivial relation can be derived for the trapped Fermi gas using Thomas-Fermi theory (see \App{thomas-fermi} for details).
Particularly, from \Eq{thomas-fermi_trapped_energy} one finds for the trapped case:
\begin{eqnarray}
\lim_{Q\to\infty} \frac{C_Q L_0}{Q^{3/2}} = \frac{8 \sqrt{2}}{9 \xi^{1/4}} \zeta\ .
\label{eq:untrapped_zeta}
\end{eqnarray}
Plugging in estimates for the trapped contact and Bertsch parameter quoted in \Tab{many-body_trapped_extrap} yields $\zeta = 0.535(2)$, which is fully consistent with the untrapped result.
Interestingly, one may also combine the results of \Eq{trapped_zeta} and \Eq{untrapped_zeta} by equating $\zeta$ in each formula to obtain a third determination of the Bertsch parameter which depends solely on the estimates of the contact for each system.
Doing so yields the value $\xi=0.372(8)$, which is consistent with the other determinations based on estimates of the energy.

\section{Conclusion}
\label{sec:conclusion}

I have performed lattice Monte Carlo studies of four-component fermion systems confined to a finite box and a harmonic trap in one spatial dimension.
I presented numerical estimates of the energies and integrated contact densities for both few- and many-body systems in the unitary limit.
The techniques used for this study relied upon a recently developed fermion world-line representation for the canonical partition function.
The main advantage of this representation is that it is free of sign problems for both polarized and unpolarized systems.
Although not considered here, the representation is also free of sign problems when there is a mass-imbalance.

It was demonstrated in \cite{Nishida:2009pg} that the unitary four-component gas at zero temperature has physical properties that are qualitatively identical to spin-1/2 fermions at unitarity in three dimensions.
Numerical studies of the one-dimensional system might therefore provide new qualitative and perhaps even quantitative insights into the nature of such nonrelativistic conformal field theories.
The main findings of this study are:
\begin{enumerate}
\item a less than one percent-level determination of continuum few-body observables for up to eight fermions confined to a harmonic trap, providing indirect estimates of the scaling dimensions of few-body operators based upon the operator state-correspondence;
\item two independent determinations of the Bertsch parameter $\xi$ to less than one percent statistical uncertainties based on estimates of the the continuum ground-state energies for trapped and untrapped many-body systems;
\item a third determination of the Bertsch parameter to within about two percent statistical uncertainties based on estimates of the associated integrated contact densities and theoretical input from a calculation based on Thomas-Fermi theory;
\item two independent determinations of the parameter $\zeta$ to within about a half percent statistical uncertainties from estimates of the integrated contact densities for trapped and untrapped many-body systems;
\item and finally, verification of the restoration of the virial theorems for all systems of fixed fermion number considered.
\end{enumerate}

Perhaps the most surprising finding of this study is an apparent numerical equivalence of Bertsch parameters for the one- and three-dimensional unitary Fermi gases.
This observation and its implications were originally reported in \cite{PhysRevLett.109.250403}.
Presently there are no known theoretical arguments for why these parameters should be equal.
Providing the equality is not by chance, one might naturally expect that conformal symmetry and scale invariance plays a crucial role in explaining the result.
Should a duality between the one- and three- dimensional systems be rigorously established in the unitary limit, then one might envision the existence of a simple prescription for relating other observables between the two theories.
Unfortunately, in the case of the contact, it remains an open question what that prescription might be.

One of the main deficiencies in the analysis presented in \Sec{analysis_and_results} is that although the finite volume scaling of dimensionless observables is well-understood from analysis of the Symanzik action, presently there is no theoretical understanding of how continuum many-body observables depend on the fermion number away from the thermodynamic limit for this system.
Consequently there is an inherent unquantifiable systematic error associated with the thermodynamic limit extrapolations of many-body continuum observables.
However, the good agreement in $\xi$ and $\zeta$ obtained from independent untrapped and trapped studies, as well as the reasonable goodness-of-fits, provide some confidence that the ansatz fit functions used for the thermodynamic limit extrapolation are reliable.
In the case of the Bertsch parameter, results are further supported by a third consistent estimate obtained by combining estimates of the contact for each system and additional theoretical input.
Nevertheless, a theoretical understanding of the scaling with $1/Q$ is highly desirable and an obvious place to start for improving the study.

Generally speaking, local updating schemes such as the one used in this study suffer from critical slowing.
It would be interesting to explore whether a worm algorithm, or continuous-time Monte Carlo approach could be applied to the nonrelativistic fermion world-line formulation in order to  improve the efficiency of the simulations.
Doing so might allow numerical simulations far closer to the continuum and infinite volume limits, and would be an important step toward achieving a high-precision (sub-percent level) determination of the Bertsch parameter.
As previously discussed, such precision estimates could have important implications for the three-dimensional unitary Fermi gas as well.

\begin{acknowledgments}
The author would like to thank J.-W. Chen, D.B. Kaplan, Y. Nishida, D.T. Son and H. Suzuki for interesting and helpful discussions.
Numerical simulations were conducted on the RIKEN Integrated Cluster of Clusters (RICC) and computer resources provided by the Theoretical High Energy Physics group at Columbia University and the RIKEN BNL Research Center.
The author is supported by the Foreign Postdoctoral Researcher program at RIKEN and by MEXT Grant-in-Aid for Young Scientists (B) (23740227).
\end{acknowledgments}

\bibliography{long}

\begin{thebibliography}{36}%
\makeatletter
\providecommand \@ifxundefined [1]{%
 \@ifx{#1\undefined}
}%
\providecommand \@ifnum [1]{%
 \ifnum #1\expandafter \@firstoftwo
 \else \expandafter \@secondoftwo
 \fi
}%
\providecommand \@ifx [1]{%
 \ifx #1\expandafter \@firstoftwo
 \else \expandafter \@secondoftwo
 \fi
}%
\providecommand \natexlab [1]{#1}%
\providecommand \enquote  [1]{``#1''}%
\providecommand \bibnamefont  [1]{#1}%
\providecommand \bibfnamefont [1]{#1}%
\providecommand \citenamefont [1]{#1}%
\providecommand \href@noop [0]{\@secondoftwo}%
\providecommand \href [0]{\begingroup \@sanitize@url \@href}%
\providecommand \@href[1]{\@@startlink{#1}\@@href}%
\providecommand \@@href[1]{\endgroup#1\@@endlink}%
\providecommand \@sanitize@url [0]{\catcode `\\12\catcode `\$12\catcode
  `\&12\catcode `\#12\catcode `\^12\catcode `\_12\catcode `\%12\relax}%
\providecommand \@@startlink[1]{}%
\providecommand \@@endlink[0]{}%
\providecommand \url  [0]{\begingroup\@sanitize@url \@url }%
\providecommand \@url [1]{\endgroup\@href {#1}{\urlprefix }}%
\providecommand \urlprefix  [0]{URL }%
\providecommand \Eprint [0]{\href }%
\providecommand \doibase [0]{http://dx.doi.org/}%
\providecommand \selectlanguage [0]{\@gobble}%
\providecommand \bibinfo  [0]{\@secondoftwo}%
\providecommand \bibfield  [0]{\@secondoftwo}%
\providecommand \translation [1]{[#1]}%
\providecommand \BibitemOpen [0]{}%
\providecommand \bibitemStop [0]{}%
\providecommand \bibitemNoStop [0]{.\EOS\space}%
\providecommand \EOS [0]{\spacefactor3000\relax}%
\providecommand \BibitemShut  [1]{\csname bibitem#1\endcsname}%
\let\auto@bib@innerbib\@empty
\bibitem [{\citenamefont {{Pethick}}\ and\ \citenamefont
  {{Ravenhall}}(1995)}]{1995ARNPS..45..429P}%
  \BibitemOpen
  \bibfield  {author} {\bibinfo {author} {\bibfnamefont {C.~J.}\ \bibnamefont
  {{Pethick}}}\ and\ \bibinfo {author} {\bibfnamefont {D.~G.}\ \bibnamefont
  {{Ravenhall}}},\ }\href {\doibase 10.1146/annurev.ns.45.120195.002241}
  {\bibfield  {journal} {\bibinfo  {journal} {Annual Review of Nuclear and
  Particle Science}\ }\textbf {\bibinfo {volume} {45}},\ \bibinfo {pages} {429}
  (\bibinfo {year} {1995})}\BibitemShut {NoStop}%
\bibitem [{\citenamefont {O'Hara}\ \emph {et~al.}(2002)\citenamefont {O'Hara},
  \citenamefont {Hemmer}, \citenamefont {Gehm}, \citenamefont {Granade},\ and\
  \citenamefont {Thomas}}]{O'Hara13122002}%
  \BibitemOpen
  \bibfield  {author} {\bibinfo {author} {\bibfnamefont {K.~M.}\ \bibnamefont
  {O'Hara}}, \bibinfo {author} {\bibfnamefont {S.~L.}\ \bibnamefont {Hemmer}},
  \bibinfo {author} {\bibfnamefont {M.~E.}\ \bibnamefont {Gehm}}, \bibinfo
  {author} {\bibfnamefont {S.~R.}\ \bibnamefont {Granade}}, \ and\ \bibinfo
  {author} {\bibfnamefont {J.~E.}\ \bibnamefont {Thomas}},\ }\href {\doibase
  10.1126/science.1079107} {\bibfield  {journal} {\bibinfo  {journal}
  {Science}\ }\textbf {\bibinfo {volume} {298}},\ \bibinfo {pages} {2179}
  (\bibinfo {year} {2002})}\BibitemShut {NoStop}%
\bibitem [{\citenamefont {Bourdel}\ \emph {et~al.}(2004)\citenamefont
  {Bourdel}, \citenamefont {Khaykovich}, \citenamefont {Cubizolles},
  \citenamefont {Zhang}, \citenamefont {Chevy}, \citenamefont {Teichmann},
  \citenamefont {Tarruell}, \citenamefont {Kokkelmans},\ and\ \citenamefont
  {Salomon}}]{PhysRevLett.93.050401}%
  \BibitemOpen
  \bibfield  {author} {\bibinfo {author} {\bibfnamefont {T.}~\bibnamefont
  {Bourdel}}, \bibinfo {author} {\bibfnamefont {L.}~\bibnamefont {Khaykovich}},
  \bibinfo {author} {\bibfnamefont {J.}~\bibnamefont {Cubizolles}}, \bibinfo
  {author} {\bibfnamefont {J.}~\bibnamefont {Zhang}}, \bibinfo {author}
  {\bibfnamefont {F.}~\bibnamefont {Chevy}}, \bibinfo {author} {\bibfnamefont
  {M.}~\bibnamefont {Teichmann}}, \bibinfo {author} {\bibfnamefont
  {L.}~\bibnamefont {Tarruell}}, \bibinfo {author} {\bibfnamefont {S.~J. J.
  M.~F.}\ \bibnamefont {Kokkelmans}}, \ and\ \bibinfo {author} {\bibfnamefont
  {C.}~\bibnamefont {Salomon}},\ }\href {\doibase
  10.1103/PhysRevLett.93.050401} {\bibfield  {journal} {\bibinfo  {journal}
  {Phys. Rev. Lett.}\ }\textbf {\bibinfo {volume} {93}},\ \bibinfo {pages}
  {050401} (\bibinfo {year} {2004})}\BibitemShut {NoStop}%
\bibitem [{\citenamefont {{Regal}}\ \emph {et~al.}(2003)\citenamefont
  {{Regal}}, \citenamefont {{Ticknor}}, \citenamefont {{Bohn}},\ and\
  \citenamefont {{Jin}}}]{2003Natur.424...47R}%
  \BibitemOpen
  \bibfield  {author} {\bibinfo {author} {\bibfnamefont {C.~A.}\ \bibnamefont
  {{Regal}}}, \bibinfo {author} {\bibfnamefont {C.}~\bibnamefont {{Ticknor}}},
  \bibinfo {author} {\bibfnamefont {J.~L.}\ \bibnamefont {{Bohn}}}, \ and\
  \bibinfo {author} {\bibfnamefont {D.~S.}\ \bibnamefont {{Jin}}},\ }\href
  {\doibase 10.1038/nature01738} {\bibfield  {journal} {\bibinfo  {journal}
  {\nat}\ }\textbf {\bibinfo {volume} {424}},\ \bibinfo {pages} {47} (\bibinfo
  {year} {2003})},\ \Eprint {http://arxiv.org/abs/arXiv:cond-mat/0305028}
  {arXiv:cond-mat/0305028} \BibitemShut {NoStop}%
\bibitem [{\citenamefont {Strecker}\ \emph {et~al.}(2003)\citenamefont
  {Strecker}, \citenamefont {Partridge},\ and\ \citenamefont
  {Hulet}}]{PhysRevLett.91.080406}%
  \BibitemOpen
  \bibfield  {author} {\bibinfo {author} {\bibfnamefont {K.~E.}\ \bibnamefont
  {Strecker}}, \bibinfo {author} {\bibfnamefont {G.~B.}\ \bibnamefont
  {Partridge}}, \ and\ \bibinfo {author} {\bibfnamefont {R.~G.}\ \bibnamefont
  {Hulet}},\ }\href {\doibase 10.1103/PhysRevLett.91.080406} {\bibfield
  {journal} {\bibinfo  {journal} {Phys. Rev. Lett.}\ }\textbf {\bibinfo
  {volume} {91}},\ \bibinfo {pages} {080406} (\bibinfo {year}
  {2003})}\BibitemShut {NoStop}%
\bibitem [{\citenamefont {Dieckmann}\ \emph {et~al.}(2002)\citenamefont
  {Dieckmann}, \citenamefont {Stan}, \citenamefont {Gupta}, \citenamefont
  {Hadzibabic}, \citenamefont {Schunck},\ and\ \citenamefont
  {Ketterle}}]{PhysRevLett.89.203201}%
  \BibitemOpen
  \bibfield  {author} {\bibinfo {author} {\bibfnamefont {K.}~\bibnamefont
  {Dieckmann}}, \bibinfo {author} {\bibfnamefont {C.~A.}\ \bibnamefont {Stan}},
  \bibinfo {author} {\bibfnamefont {S.}~\bibnamefont {Gupta}}, \bibinfo
  {author} {\bibfnamefont {Z.}~\bibnamefont {Hadzibabic}}, \bibinfo {author}
  {\bibfnamefont {C.~H.}\ \bibnamefont {Schunck}}, \ and\ \bibinfo {author}
  {\bibfnamefont {W.}~\bibnamefont {Ketterle}},\ }\href {\doibase
  10.1103/PhysRevLett.89.203201} {\bibfield  {journal} {\bibinfo  {journal}
  {Phys. Rev. Lett.}\ }\textbf {\bibinfo {volume} {89}},\ \bibinfo {pages}
  {203201} (\bibinfo {year} {2002})}\BibitemShut {NoStop}%
\bibitem [{\citenamefont {Baker}(2000)}]{baker2000mbx}%
  \BibitemOpen
  \bibfield  {author} {\bibinfo {author} {\bibfnamefont {G.}~\bibnamefont
  {Baker}},\ }in\ \href@noop {} {\emph {\bibinfo {booktitle} {Recent progress
  in many-body theories: the proceedings of the 10th international conference,
  Seattle, USA, September 10-15, 1999}}},\ Vol.~\bibinfo {volume} {3}\
  (\bibinfo {organization} {World Scientific Pub Co Inc},\ \bibinfo {year}
  {2000})\ p.~\bibinfo {pages} {15}\BibitemShut {NoStop}%
\bibitem [{\citenamefont {Papenbrock}(2005)}]{PhysRevA.72.041603}%
  \BibitemOpen
  \bibfield  {author} {\bibinfo {author} {\bibfnamefont {T.}~\bibnamefont
  {Papenbrock}},\ }\href {\doibase 10.1103/PhysRevA.72.041603} {\bibfield
  {journal} {\bibinfo  {journal} {Phys. Rev. A}\ }\textbf {\bibinfo {volume}
  {72}},\ \bibinfo {pages} {041603} (\bibinfo {year} {2005})}\BibitemShut
  {NoStop}%
\bibitem [{\citenamefont {Son}\ and\ \citenamefont
  {Wingate}(2006)}]{Son:2005rv}%
  \BibitemOpen
  \bibfield  {author} {\bibinfo {author} {\bibfnamefont {D.}~\bibnamefont
  {Son}}\ and\ \bibinfo {author} {\bibfnamefont {M.}~\bibnamefont {Wingate}},\
  }\href {\doibase 10.1016/j.aop.2005.11.001} {\bibfield  {journal} {\bibinfo
  {journal} {Annals Phys.}\ }\textbf {\bibinfo {volume} {321}},\ \bibinfo
  {pages} {197} (\bibinfo {year} {2006})},\ \Eprint
  {http://arxiv.org/abs/cond-mat/0509786} {arXiv:cond-mat/0509786 [cond-mat]}
  \BibitemShut {NoStop}%
\bibitem [{\citenamefont {{Ma{\~n}es}}\ and\ \citenamefont
  {{Valle}}(2009)}]{2009AnPhy.324.1136M}%
  \BibitemOpen
  \bibfield  {author} {\bibinfo {author} {\bibfnamefont {J.~L.}\ \bibnamefont
  {{Ma{\~n}es}}}\ and\ \bibinfo {author} {\bibfnamefont {M.~A.}\ \bibnamefont
  {{Valle}}},\ }\href {\doibase 10.1016/j.aop.2009.01.003} {\bibfield
  {journal} {\bibinfo  {journal} {Annals of Physics}\ }\textbf {\bibinfo
  {volume} {324}},\ \bibinfo {pages} {1136} (\bibinfo {year} {2009})},\ \Eprint
  {http://arxiv.org/abs/0810.3797} {arXiv:0810.3797 [cond-mat.other]}
  \BibitemShut {NoStop}%
\bibitem [{\citenamefont {Endres}\ \emph {et~al.}(2013)\citenamefont {Endres},
  \citenamefont {Kaplan}, \citenamefont {Lee},\ and\ \citenamefont
  {Nicholson}}]{PhysRevA.87.023615}%
  \BibitemOpen
  \bibfield  {author} {\bibinfo {author} {\bibfnamefont {M.~G.}\ \bibnamefont
  {Endres}}, \bibinfo {author} {\bibfnamefont {D.~B.}\ \bibnamefont {Kaplan}},
  \bibinfo {author} {\bibfnamefont {J.-W.}\ \bibnamefont {Lee}}, \ and\
  \bibinfo {author} {\bibfnamefont {A.~N.}\ \bibnamefont {Nicholson}},\ }\href
  {\doibase 10.1103/PhysRevA.87.023615} {\bibfield  {journal} {\bibinfo
  {journal} {Phys. Rev. A}\ }\textbf {\bibinfo {volume} {87}},\ \bibinfo
  {pages} {023615} (\bibinfo {year} {2013})}\BibitemShut {NoStop}%
\bibitem [{\citenamefont {Carlson}\ \emph {et~al.}(2011)\citenamefont
  {Carlson}, \citenamefont {Gandolfi}, \citenamefont {Schmidt},\ and\
  \citenamefont {Zhang}}]{PhysRevA.84.061602}%
  \BibitemOpen
  \bibfield  {author} {\bibinfo {author} {\bibfnamefont {J.}~\bibnamefont
  {Carlson}}, \bibinfo {author} {\bibfnamefont {S.}~\bibnamefont {Gandolfi}},
  \bibinfo {author} {\bibfnamefont {K.~E.}\ \bibnamefont {Schmidt}}, \ and\
  \bibinfo {author} {\bibfnamefont {S.}~\bibnamefont {Zhang}},\ }\href
  {\doibase 10.1103/PhysRevA.84.061602} {\bibfield  {journal} {\bibinfo
  {journal} {Phys. Rev. A}\ }\textbf {\bibinfo {volume} {84}},\ \bibinfo
  {pages} {061602} (\bibinfo {year} {2011})}\BibitemShut {NoStop}%
\bibitem [{\citenamefont {Ku}\ \emph {et~al.}(2012)\citenamefont {Ku},
  \citenamefont {Sommer}, \citenamefont {Cheuk},\ and\ \citenamefont
  {Zwierlein}}]{Ku03022012}%
  \BibitemOpen
  \bibfield  {author} {\bibinfo {author} {\bibfnamefont {M.~J.~H.}\
  \bibnamefont {Ku}}, \bibinfo {author} {\bibfnamefont {A.~T.}\ \bibnamefont
  {Sommer}}, \bibinfo {author} {\bibfnamefont {L.~W.}\ \bibnamefont {Cheuk}}, \
  and\ \bibinfo {author} {\bibfnamefont {M.~W.}\ \bibnamefont {Zwierlein}},\
  }\href {\doibase 10.1126/science.1214987} {\bibfield  {journal} {\bibinfo
  {journal} {Science}\ }\textbf {\bibinfo {volume} {335}},\ \bibinfo {pages}
  {563} (\bibinfo {year} {2012})}\BibitemShut {NoStop}%
\bibitem [{\citenamefont {Nishida}\ and\ \citenamefont
  {Tan}(2008)}]{PhysRevLett.101.170401}%
  \BibitemOpen
  \bibfield  {author} {\bibinfo {author} {\bibfnamefont {Y.}~\bibnamefont
  {Nishida}}\ and\ \bibinfo {author} {\bibfnamefont {S.}~\bibnamefont {Tan}},\
  }\href {\doibase 10.1103/PhysRevLett.101.170401} {\bibfield  {journal}
  {\bibinfo  {journal} {Phys. Rev. Lett.}\ }\textbf {\bibinfo {volume} {101}},\
  \bibinfo {pages} {170401} (\bibinfo {year} {2008})}\BibitemShut {NoStop}%
\bibitem [{\citenamefont {Nishida}\ and\ \citenamefont
  {Son}(2010)}]{Nishida:2009pg}%
  \BibitemOpen
  \bibfield  {author} {\bibinfo {author} {\bibfnamefont {Y.}~\bibnamefont
  {Nishida}}\ and\ \bibinfo {author} {\bibfnamefont {D.~T.}\ \bibnamefont
  {Son}},\ }\href {\doibase 10.1103/PhysRevA.82.043606} {\bibfield  {journal}
  {\bibinfo  {journal} {Phys.Rev.}\ }\textbf {\bibinfo {volume} {A82}},\
  \bibinfo {pages} {043606} (\bibinfo {year} {2010})},\ \Eprint
  {http://arxiv.org/abs/0908.2159} {arXiv:0908.2159 [cond-mat.quant-gas]}
  \BibitemShut {NoStop}%
\bibitem [{\citenamefont {Endres}(2012)}]{PhysRevLett.109.250403}%
  \BibitemOpen
  \bibfield  {author} {\bibinfo {author} {\bibfnamefont {M.~G.}\ \bibnamefont
  {Endres}},\ }\href {\doibase 10.1103/PhysRevLett.109.250403} {\bibfield
  {journal} {\bibinfo  {journal} {Phys. Rev. Lett.}\ }\textbf {\bibinfo
  {volume} {109}},\ \bibinfo {pages} {250403} (\bibinfo {year}
  {2012})}\BibitemShut {NoStop}%
\bibitem [{\citenamefont {{Tan}}(2008{\natexlab{a}})}]{2008AnPhy.323.2952T}%
  \BibitemOpen
  \bibfield  {author} {\bibinfo {author} {\bibfnamefont {S.}~\bibnamefont
  {{Tan}}},\ }\href {\doibase 10.1016/j.aop.2008.03.004} {\bibfield  {journal}
  {\bibinfo  {journal} {Annals of Physics}\ }\textbf {\bibinfo {volume}
  {323}},\ \bibinfo {pages} {2952} (\bibinfo {year} {2008}{\natexlab{a}})},\
  \Eprint {http://arxiv.org/abs/cond-mat/0505200} {cond-mat/0505200}
  \BibitemShut {NoStop}%
\bibitem [{\citenamefont {{Tan}}(2008{\natexlab{b}})}]{2008AnPhy.323.2971T}%
  \BibitemOpen
  \bibfield  {author} {\bibinfo {author} {\bibfnamefont {S.}~\bibnamefont
  {{Tan}}},\ }\href {\doibase 10.1016/j.aop.2008.03.005} {\bibfield  {journal}
  {\bibinfo  {journal} {Annals of Physics}\ }\textbf {\bibinfo {volume}
  {323}},\ \bibinfo {pages} {2971} (\bibinfo {year} {2008}{\natexlab{b}})},\
  \Eprint {http://arxiv.org/abs/cond-mat/0508320} {cond-mat/0508320}
  \BibitemShut {NoStop}%
\bibitem [{\citenamefont {{Tan}}(2008{\natexlab{c}})}]{2008AnPhy.323.2987T}%
  \BibitemOpen
  \bibfield  {author} {\bibinfo {author} {\bibfnamefont {S.}~\bibnamefont
  {{Tan}}},\ }\href {\doibase 10.1016/j.aop.2008.03.003} {\bibfield  {journal}
  {\bibinfo  {journal} {Annals of Physics}\ }\textbf {\bibinfo {volume}
  {323}},\ \bibinfo {pages} {2987} (\bibinfo {year} {2008}{\natexlab{c}})},\
  \Eprint {http://arxiv.org/abs/arXiv:0803.0841} {arXiv:arXiv:0803.0841
  [cond-mat.stat-mech]} \BibitemShut {NoStop}%
\bibitem [{\citenamefont {Braaten}\ and\ \citenamefont
  {Platter}(2008)}]{PhysRevLett.100.205301}%
  \BibitemOpen
  \bibfield  {author} {\bibinfo {author} {\bibfnamefont {E.}~\bibnamefont
  {Braaten}}\ and\ \bibinfo {author} {\bibfnamefont {L.}~\bibnamefont
  {Platter}},\ }\href {\doibase 10.1103/PhysRevLett.100.205301} {\bibfield
  {journal} {\bibinfo  {journal} {Phys. Rev. Lett.}\ }\textbf {\bibinfo
  {volume} {100}},\ \bibinfo {pages} {205301} (\bibinfo {year}
  {2008})}\BibitemShut {NoStop}%
\bibitem [{\citenamefont {{Braaten}}\ \emph {et~al.}(2008)\citenamefont
  {{Braaten}}, \citenamefont {{Kang}},\ and\ \citenamefont
  {{Platter}}}]{2008PhRvA..78e3606B}%
  \BibitemOpen
  \bibfield  {author} {\bibinfo {author} {\bibfnamefont {E.}~\bibnamefont
  {{Braaten}}}, \bibinfo {author} {\bibfnamefont {D.}~\bibnamefont {{Kang}}}, \
  and\ \bibinfo {author} {\bibfnamefont {L.}~\bibnamefont {{Platter}}},\ }\href
  {\doibase 10.1103/PhysRevA.78.053606} {\bibfield  {journal} {\bibinfo
  {journal} {\pra}\ }\textbf {\bibinfo {volume} {78}},\ \bibinfo {eid} {053606}
  (\bibinfo {year} {2008})},\ \Eprint {http://arxiv.org/abs/0806.2277}
  {arXiv:0806.2277 [cond-mat.other]} \BibitemShut {NoStop}%
\bibitem [{\citenamefont {Hagen}(1972)}]{PhysRevD.5.377}%
  \BibitemOpen
  \bibfield  {author} {\bibinfo {author} {\bibfnamefont {C.~R.}\ \bibnamefont
  {Hagen}},\ }\href {\doibase 10.1103/PhysRevD.5.377} {\bibfield  {journal}
  {\bibinfo  {journal} {Phys. Rev. D}\ }\textbf {\bibinfo {volume} {5}},\
  \bibinfo {pages} {377} (\bibinfo {year} {1972})}\BibitemShut {NoStop}%
\bibitem [{\citenamefont {Niederer}(1972)}]{Niederer:1972zz}%
  \BibitemOpen
  \bibfield  {author} {\bibinfo {author} {\bibfnamefont {U.}~\bibnamefont
  {Niederer}},\ }\href@noop {} {\bibfield  {journal} {\bibinfo  {journal}
  {Helv.Phys.Acta}\ }\textbf {\bibinfo {volume} {45}},\ \bibinfo {pages} {802}
  (\bibinfo {year} {1972})}\BibitemShut {NoStop}%
\bibitem [{\citenamefont {Nishida}\ and\ \citenamefont
  {Son}(2007)}]{PhysRevD.76.086004}%
  \BibitemOpen
  \bibfield  {author} {\bibinfo {author} {\bibfnamefont {Y.}~\bibnamefont
  {Nishida}}\ and\ \bibinfo {author} {\bibfnamefont {D.~T.}\ \bibnamefont
  {Son}},\ }\href {\doibase 10.1103/PhysRevD.76.086004} {\bibfield  {journal}
  {\bibinfo  {journal} {Phys. Rev. D}\ }\textbf {\bibinfo {volume} {76}},\
  \bibinfo {pages} {086004} (\bibinfo {year} {2007})}\BibitemShut {NoStop}%
\bibitem [{\citenamefont {Mehen}(2008)}]{PhysRevA.78.013614}%
  \BibitemOpen
  \bibfield  {author} {\bibinfo {author} {\bibfnamefont {T.}~\bibnamefont
  {Mehen}},\ }\href {\doibase 10.1103/PhysRevA.78.013614} {\bibfield  {journal}
  {\bibinfo  {journal} {Phys. Rev. A}\ }\textbf {\bibinfo {volume} {78}},\
  \bibinfo {pages} {013614} (\bibinfo {year} {2008})}\BibitemShut {NoStop}%
\bibitem [{\citenamefont {Chen}\ and\ \citenamefont
  {Kaplan}(2004)}]{PhysRevLett.92.257002}%
  \BibitemOpen
  \bibfield  {author} {\bibinfo {author} {\bibfnamefont {J.-W.}\ \bibnamefont
  {Chen}}\ and\ \bibinfo {author} {\bibfnamefont {D.~B.}\ \bibnamefont
  {Kaplan}},\ }\href {\doibase 10.1103/PhysRevLett.92.257002} {\bibfield
  {journal} {\bibinfo  {journal} {Phys. Rev. Lett.}\ }\textbf {\bibinfo
  {volume} {92}},\ \bibinfo {pages} {257002} (\bibinfo {year}
  {2004})}\BibitemShut {NoStop}%
\bibitem [{\citenamefont {{Kaplan}}\ \emph {et~al.}(1996)\citenamefont
  {{Kaplan}}, \citenamefont {{Savage}},\ and\ \citenamefont
  {{Wise}}}]{1996NuPhB.478..629K}%
  \BibitemOpen
  \bibfield  {author} {\bibinfo {author} {\bibfnamefont {D.~B.}\ \bibnamefont
  {{Kaplan}}}, \bibinfo {author} {\bibfnamefont {M.~J.}\ \bibnamefont
  {{Savage}}}, \ and\ \bibinfo {author} {\bibfnamefont {M.~B.}\ \bibnamefont
  {{Wise}}},\ }\href {\doibase 10.1016/0550-3213(96)00357-4} {\bibfield
  {journal} {\bibinfo  {journal} {Nuclear Physics B}\ }\textbf {\bibinfo
  {volume} {478}},\ \bibinfo {pages} {629} (\bibinfo {year} {1996})},\ \Eprint
  {http://arxiv.org/abs/arXiv:nucl-th/9605002} {arXiv:nucl-th/9605002}
  \BibitemShut {NoStop}%
\bibitem [{\citenamefont {{Stratonovich}}(1957)}]{1957SPhD....2..416S}%
  \BibitemOpen
  \bibfield  {author} {\bibinfo {author} {\bibfnamefont {R.~L.}\ \bibnamefont
  {{Stratonovich}}},\ }\href@noop {} {\bibfield  {journal} {\bibinfo  {journal}
  {Soviet Physics Doklady}\ }\textbf {\bibinfo {volume} {2}},\ \bibinfo {pages}
  {416} (\bibinfo {year} {1957})}\BibitemShut {NoStop}%
\bibitem [{\citenamefont {Hubbard}(1959)}]{PhysRevLett.3.77}%
  \BibitemOpen
  \bibfield  {author} {\bibinfo {author} {\bibfnamefont {J.}~\bibnamefont
  {Hubbard}},\ }\href {\doibase 10.1103/PhysRevLett.3.77} {\bibfield  {journal}
  {\bibinfo  {journal} {Phys. Rev. Lett.}\ }\textbf {\bibinfo {volume} {3}},\
  \bibinfo {pages} {77} (\bibinfo {year} {1959})}\BibitemShut {NoStop}%
\bibitem [{\citenamefont {{Endres}}(2012)}]{2012PhRvA..85f3624E}%
  \BibitemOpen
  \bibfield  {author} {\bibinfo {author} {\bibfnamefont {M.~G.}\ \bibnamefont
  {{Endres}}},\ }\href {\doibase 10.1103/PhysRevA.85.063624} {\bibfield
  {journal} {\bibinfo  {journal} {\pra}\ }\textbf {\bibinfo {volume} {85}},\
  \bibinfo {eid} {063624} (\bibinfo {year} {2012})},\ \Eprint
  {http://arxiv.org/abs/1204.6182} {arXiv:1204.6182 [hep-lat]} \BibitemShut
  {NoStop}%
\bibitem [{\citenamefont {Itzykson}\ and\ \citenamefont
  {Drouffe}(1989)}]{itzykson:book}%
  \BibitemOpen
  \bibfield  {author} {\bibinfo {author} {\bibfnamefont {C.}~\bibnamefont
  {Itzykson}}\ and\ \bibinfo {author} {\bibfnamefont {J.~M.}\ \bibnamefont
  {Drouffe}},\ }\href@noop {} {\emph {\bibinfo {title} {{Statistical Field
  Theory}}}}\ (\bibinfo  {publisher} {Cambridge University Press},\ \bibinfo
  {address} {Cambridge},\ \bibinfo {year} {1989})\BibitemShut {NoStop}%
\bibitem [{\citenamefont {Luscher}(1994)}]{Luscher1994100}%
  \BibitemOpen
  \bibfield  {author} {\bibinfo {author} {\bibfnamefont {M.}~\bibnamefont
  {Luscher}},\ }\href {\doibase 10.1016/0010-4655(94)90232-1} {\bibfield
  {journal} {\bibinfo  {journal} {Computer Physics Communications}\ }\textbf
  {\bibinfo {volume} {79}},\ \bibinfo {pages} {100 } (\bibinfo {year}
  {1994})}\BibitemShut {NoStop}%
\bibitem [{\citenamefont {Endres}\ \emph {et~al.}(2011)\citenamefont {Endres},
  \citenamefont {Kaplan}, \citenamefont {Lee},\ and\ \citenamefont
  {Nicholson}}]{PhysRevA.84.043644}%
  \BibitemOpen
  \bibfield  {author} {\bibinfo {author} {\bibfnamefont {M.~G.}\ \bibnamefont
  {Endres}}, \bibinfo {author} {\bibfnamefont {D.~B.}\ \bibnamefont {Kaplan}},
  \bibinfo {author} {\bibfnamefont {J.-W.}\ \bibnamefont {Lee}}, \ and\
  \bibinfo {author} {\bibfnamefont {A.~N.}\ \bibnamefont {Nicholson}},\ }\href
  {\doibase 10.1103/PhysRevA.84.043644} {\bibfield  {journal} {\bibinfo
  {journal} {Phys. Rev. A}\ }\textbf {\bibinfo {volume} {84}},\ \bibinfo
  {pages} {043644} (\bibinfo {year} {2011})}\BibitemShut {NoStop}%
\bibitem [{\citenamefont {Symanzik}(1983{\natexlab{a}})}]{Symanzik1983187}%
  \BibitemOpen
  \bibfield  {author} {\bibinfo {author} {\bibfnamefont {K.}~\bibnamefont
  {Symanzik}},\ }\href {\doibase 10.1016/0550-3213(83)90468-6} {\bibfield
  {journal} {\bibinfo  {journal} {Nuclear Physics B}\ }\textbf {\bibinfo
  {volume} {226}},\ \bibinfo {pages} {187 } (\bibinfo {year}
  {1983}{\natexlab{a}})}\BibitemShut {NoStop}%
\bibitem [{\citenamefont {Symanzik}(1983{\natexlab{b}})}]{Symanzik1983205}%
  \BibitemOpen
  \bibfield  {author} {\bibinfo {author} {\bibfnamefont {K.}~\bibnamefont
  {Symanzik}},\ }\href {\doibase 10.1016/0550-3213(83)90469-8} {\bibfield
  {journal} {\bibinfo  {journal} {Nuclear Physics B}\ }\textbf {\bibinfo
  {volume} {226}},\ \bibinfo {pages} {205 } (\bibinfo {year}
  {1983}{\natexlab{b}})}\BibitemShut {NoStop}%
\bibitem [{sup()}]{supp}%
  \BibitemOpen
  \href@noop {} {\ }\bibinfo {note} {See Supplemental Material at
  http://link.aps.org/supplemental/10.1103/PhysRevA.87.063617 for fit results
  for various physical observables.}\BibitemShut {Stop}%
\end{thebibliography}%

\appendix

\section{Trapped many-body energies from Thomas-Fermi theory}
\label{app:thomas-fermi}

Here, I briefly derive \Eq{trapped_energy} and its subleading correction in $1/a$ using a simple density-functional theory calculation following \cite{PhysRevA.72.041603}.
The Thomas-Fermi density functional for harmonically trapped fermions is given by
\begin{eqnarray}
E^{osc}_{TF}[\rho] = \int dx \, \left[ \calE(\rho(x)) + \rho(x) v(x)  \right]\ ,
\label{eq:thomas-fermi_energy}
\end{eqnarray}
where
\begin{eqnarray}
\calE(\rho) =  \calE_0(\rho) \left( \xi - \frac{\zeta}{k_F a} + \ldots \right)
\end{eqnarray}
is the energy per unit volume of the system with $\calE_0(\rho)$ defined in \Eq{free_gas_energy}, and $v(x)$ is the external harmonic trapping potential defined in \Eq{potential}.
The task is to minimize $E^{osc}_{TF}[\rho]$ with respect to $\rho(x)$, subject to the constraint that the total number of fermions 
\begin{eqnarray}
Q = \int dx \, \rho(x)
\label{eq:thomas-fermi_charge}
\end{eqnarray}
is held fixed.
Introducing a Lagrange multiplier $\mu$ (i.e., a chemical potential) to enforce \Eq{thomas-fermi_charge} as a constraint, one finds that the functional is extremized by solutions to
\begin{eqnarray}
\rho^2 - \frac{8\zeta}{3\pi\xi a} \rho = \rho_0^2 \left( 1 - \frac{x^2}{x_0^2}\right)
\label{eq:thomas-fermi_min}
\end{eqnarray}
with
\begin{eqnarray}
\rho_0 = \sqrt{\frac{32 m \mu}{\xi \pi^2} }\ ,\qquad x_0 = \sqrt{\frac{2\mu}{m\omega^2}}\ .
\label{eq:thomas-fermi_defs}
\end{eqnarray}

Solving \Eq{thomas-fermi_min} perturbatively in $(\rho_0 a)^{-1}$ yields the solution
\begin{eqnarray}
\rho(x) = \rho_0 \left( 1 - \frac{x^2}{x_0^2}  \right)^{1/2} + \frac{4 \zeta}{3\pi \xi a} + \ldots
\label{eq:thomas-fermi_density}
\end{eqnarray}
up to corrections of order $\calO(\rho_0 a)^{-2}$.
Combining \Eq{thomas-fermi_charge} with \Eq{thomas-fermi_density}, one may relate the chemical potential to the charge, finding
\begin{eqnarray}
Q = \frac{4\mu}{\sqrt{\xi}\omega} + \frac{8\zeta}{3\pi\xi a } \sqrt{ \frac{2\mu}{m \omega^2} } + \ldots 
\end{eqnarray}
Inverting this relation yields $\mu$ as a function of $Q$, given by:
\begin{eqnarray}
\mu = \frac{1}{4} \sqrt{\xi} Q \omega - \frac{ 1 }{3\pi a} \sqrt{\frac{2 Q \omega}{m }} \frac{ \zeta }{\xi^{1/4}} +\ldots\ .
\label{eq:thomas-fermi_mu}
\end{eqnarray}
Plugging \Eq{thomas-fermi_density} back into \Eq{thomas-fermi_energy}, and using \Eq{thomas-fermi_defs} and \Eq{thomas-fermi_mu} yields 
\begin{eqnarray}
E^{osc}(Q) = \sqrt{\xi} E^{osc}_0(Q) - \frac{2 }{9 \pi a } \sqrt{\frac{2 \omega}{m}} Q^{3/2} \frac{\zeta}{\xi^{1/4}}+\ldots  \ ,
\label{eq:thomas-fermi_trapped_energy}
\end{eqnarray}
for the energy for trapped fermions in the unitary regime, where $E^{osc}_0(Q)$ is given by \Eq{free_trapped_gas_energy}.

\end{document}